\documentclass[twocolumn]{aastex631}
\usepackage{showyourwork}
\usepackage{multirow}
\usepackage{datetime}
\yyyymmdddate
\usepackage{mathtools}
\usepackage{savesym}
\let\tablenum\relax
\usepackage{siunitx}[=v2]
\let\oldenddeluxetable\endxdeluxetable
\let\olddeluxetable\xdeluxetable
\makeatletter
\renewenvironment{xdeluxetable}[1]{
\olddeluxetable{[#1]}
\def\pt@format{\string#1}%
}{\oldenddeluxetable}
\makeatother

\DeclareSIUnit{\mag}{mag}
\DeclareSIUnit{\mas}{mas}
\DeclareSIUnit{\electron}{e^-}
\DeclareSIUnit{\pixel}{px}
\DeclareSIUnit{\adu}{adu}
\DeclareSIUnit{\year}{yr}
\DeclareSIUnit{\parsec}{pc}
\DeclareSIUnit{\au}{au}

\begin{document}

\title{Visible-Light High-Contrast Imaging and Polarimetry with SCExAO/VAMPIRES}
\shorttitle{Visible-Light High-Contrast Imaging \& Polarimetry with SCExAO/VAMPIRES}

\received{2024/08/21}
\revised{2024/10/15}
\submitjournal{PASP}

\author[0000-0001-6341-310X]{Miles Lucas}
\affiliation{Institute for Astronomy, University of Hawai'i, 640 N. Aohoku Pl., Hilo, HI 96720, USA}

\author[0000-0002-8352-7515]{Barnaby Norris}
\affiliation{Sydney Institute for Astronomy, School of Physics, Physics Road, University of Sydney, NSW 2006, Australia}
\affiliation{Sydney Astrophotonic Instrumentation Laboratories, Physics Road, University of Sydney, NSW 2006, Australia}
\affiliation{AAO-USyd, School of Physics, University of Sydney, NSW 2006, Australia}

\author[0000-0002-1097-9908]{Olivier Guyon}
\affiliation{Subaru Telescope, National Astronomical Observatory of Japan, 650 N. Aohoku Pl., Hilo, HI 96720, USA}
\affiliation{College of Optical Sciences, University of Arizona, Tucson, AZ 87521, USA}
\affiliation{Steward Observatory, University of Arizona, Tucson, AZ 87521, USA}
\affiliation{Astrobiology Center, 2 Chome-21-1, Osawa, Mitaka, Tokyo, 181-8588, Japan}

\author[0000-0003-1341-5531]{Michael Bottom}
\affiliation{Institute for Astronomy, University of Hawai'i, 640 N. Aohoku Pl., Hilo, HI 96720, USA}

\author[0000-0003-4514-7906]{Vincent Deo}
\affiliation{Subaru Telescope, National Astronomical Observatory of Japan, 650 N. Aohoku Pl., Hilo, HI 96720, USA}

\author[0000-0003-4018-2569]{S\'ebastien Vievard}
\affiliation{Subaru Telescope, National Astronomical Observatory of Japan, 650 N. Aohoku Pl., Hilo, HI 96720, USA}
\affiliation{Astrobiology Center, 2 Chome-21-1, Osawa, Mitaka, Tokyo, 181-8588, Japan}

\author[0000-0002-3047-1845]{Julien Lozi}
\affiliation{Subaru Telescope, National Astronomical Observatory of Japan, 650 N. Aohoku Pl., Hilo, HI 96720, USA}

\author[0000-0002-1094-852X]{Kyohoon Ahn}
\affiliation{Subaru Telescope, National Astronomical Observatory of Japan, 650 N. Aohoku Pl., Hilo, HI 96720, USA}
\affiliation{Korea Astronomy and Space Science Institute, 776 Daedeok-daero, Yuseong-gu, Daejeon 34055, Republic of Korea}

\author[0000-0001-5082-7442]{Jaren Ashcraft}
\affiliation{College of Optical Sciences, University of Arizona, Tucson, AZ 87521, USA}

\author[0000-0002-7405-3119]{Thayne Currie}
\affiliation{Subaru Telescope, National Astronomical Observatory of Japan, 650 N. Aohoku Pl., Hilo, HI 96720, USA}
\affiliation{Department of Physics and Astronomy, University of Texas at San Antonio, One UTSA Circle, San Antonio, TX 78249, USA}

\author[0000-0003-0695-0480]{David Doelman}
\affiliation{Leiden Observatory, Leiden University, P.O. Box 9513, 2300 RA Leiden, The Netherlands}
\affiliation{SRON Netherlands Institute for Space Research, Niels Bohrweg 4, 2333 CA, Leiden, The Netherlands}

\author[0000-0002-9294-1793]{Tomoyuki Kudo}
\affiliation{Subaru Telescope, National Astronomical Observatory of Japan, 650 N. Aohoku Pl., Hilo, HI 96720, USA}

\author[0000-0002-8566-2577]{Lucie Leboulleux}
\affiliation{University Grenoble Alpes, CNRS, IPAG, 38000 Grenoble, France}

\author[0009-0009-6274-6514]{Lucinda Lilley}
\affiliation{Sydney Institute for Astronomy, School of Physics, Physics Road, University of Sydney, NSW 2006, Australia}

\affiliation{Sydney Astrophotonic Instrumentation Laboratories, Physics Road, University of Sydney, NSW 2006, Australia}

\author[0000-0001-6205-9233]{Maxwell Millar-Blanchaer}
\affiliation{Department of Physics, University of California, Santa Barbara, CA, 93106, USA}

\author[0000-0003-1713-3208]{Boris Safonov}
\affiliation{Sternberg Astronomical Institute, Lomonosov Moscow State Univeristy, 119992 Universitetskii prospekt 13, Moscow, Russia}

\author[0000-0001-7026-6291]{Peter Tuthill}
\affiliation{Sydney Institute for Astronomy, School of Physics, Physics Road, University of Sydney, NSW 2006, Australia}

\author[0000-0002-6879-3030]{Taichi Uyama}
\affiliation{Department of Physics and Astronomy, California State University, Northridge, Northridge, CA 91330 USA}

\author[0009-0005-0321-0104]{Aidan Walk}
\affiliation{Institute for Astronomy, University of Hawai'i, 640 N. Aohoku Pl., Hilo, HI 96720, USA}
\affiliation{Subaru Telescope, National Astronomical Observatory of Japan, 650 N. Aohoku Pl., Hilo, HI 96720, USA}

\author[0000-0003-3567-6839]{Manxuan Zhang}
\affiliation{Department of Physics, University of California, Santa Barbara, CA, 93106, USA}

\begin{abstract}
    We present significant upgrades to the VAMPIRES instrument, a visible-light (\SIrange{600}{800}{\nano\meter}) high-contrast imaging polarimeter integrated within SCExAO on the Subaru telescope. Key enhancements include new qCMOS detectors, coronagraphs, polarization optics, and a multiband imaging mode, improving sensitivity, resolution, and efficiency. These upgrades position VAMPIRES as a powerful tool for studying sub-stellar companions, accreting protoplanets, circumstellar disks, stellar jets, stellar mass-loss shells, and solar system objects. The instrument achieves angular resolutions from \SIrange{17}{21}{\mas} and Strehl ratios up to 60\%, with 5$\sigma$ contrast limits of $10^{\text{-}4}$ at \ang{;;0.1} to $10^{\text{-}6}$ beyond \ang{;;0.5}. We demonstrate these capabilities through spectro-polarimetric coronagraphic imaging of the HD 169142 circumstellar disk, ADI+SDI imaging of the sub-stellar companion HD 1160B, narrowband H$\alpha$ imaging of the R Aqr emission nebula, and spectro-polarimetric imaging of Neptune.
\end{abstract}

\section{Introduction}\label{sec:intro}

The advent of \SI{8}{\meter}-class telescopes combined with the advances in extreme adaptive optics has revolutionized astronomy by enabling high-angular resolution and high-sensitivity imaging and spectroscopy \citep{guyon_extreme_2018,currie_direct_2023,galicher_imaging_2024}. By employing cutting-edge technologies, innovative observing techniques, and sophisticated post-processing methods, high-contrast imaging (HCI) mitigates the diffraction effects of the stellar point spread function (PSF), enabling the study of faint circumstellar signals.

In recent years, the majority of HCI science has occurred at near-infrared (NIR) and thermal-infrared (TIR) wavelengths (\SIrange{1}{5}{\micron}) because longer wavelengths are affected less by atmospheric turbulence \citep{fried_optical_1966,roddier_effects_1981}. Currently, HCI instruments can reach $>$90\% Strehl ratios, a measure of the PSF quality, in good conditions in the NIR \citep{beuzit_sphere_2019,lozi_status_2020}. Stable, high-quality PSFs are required for efficient diffraction control via coronagraphic or interferometric techniques.

High-contrast imaging instruments are more sensitive to optical path differences in visible light than in NIR \citep{fried_optical_1966,roddier_effects_1981}. Even the most advanced AO systems can only reach 40\%-60\% Strehl ratios at visible wavelengths \citep{ahn_scexao_2021,males_magao-x_2022}. This can be mitigated with ``lucky imaging'', where images are taken faster than the speckle coherence time ($\tau_0 \sim$\SI{4}{\milli\second}, \citealp{kooten_climate_2022}) ``freezing'' the speckle pattern, allowing individual frames with high-quality point spread functions (PSFs) to be aligned and combined while discarding poor-quality frames \citep{law_lucky_2006, garrel_highly_2012, lang_tractor_2016}.

Despite the challenges in wavefront control, visible-light instruments have some key advantages \citep{close_into_2014}.  Visible wavelengths have a smaller diffraction limit and, therefore, higher angular resolution than the NIR. The sky background is orders of magnitude fainter in the visible. Stellar emission lines in the visible (e.g., H$\alpha$, OIII, SII) have much higher contrasts compared to their local continuum than in the NIR (e.g., Pa$\beta$). Visible detectors have better noise characteristics and operational flexibility than infrared detectors, including more compact assemblies, simpler cooling requirements, and lower cost.

Specific science cases for visible HCI include circumstellar disks \citep{benisty_optical_2023}, exoplanet and sub-stellar companions \citep{hunziker_refplanets_2020}, evolved stars' atmospheres and mass-loss shells \citep{norris_vampires_2015}, stellar jets \citep{schmid_spherezimpol_2017,uyama_monitoring_2022}, accreting protoplanets \citep{uyama_high-contrast_2020,benisty_circumplanetary_2021}, close binaries \citep{mcclure_binary_1980,escorza_barium_2019}, and solar system objects \citep{schmid_limb_2006,vernazza_vltsphere_2021}.

Visible-light high contrast instruments include SCExAO/VAMPIRES \citep{norris_vampires_2015}, SPHERE/ZIMPOL \citep{schmid_spherezimpol_2018}, MagAO-X/VisAO \citep{males_magao-x_2024}, and LBT/SHARK-VIS \citep{mattioli_shark-vis_2018}. These instruments share some commonalities. They are all deployed on large telescopes with extreme AO correction-- high-order deformable mirrors and high framerate, photon-counting wavefront sensors. VisAO, SHARK-VIS, and VAMPIRES use high-framerate detectors for lucky imaging. All instruments have some form of narrowband spectral emission line imaging. Finally, VAMPIRES and ZIMPOL offer polarimetric modes, which use high-speed polarimetric modulation to combat atmospheric seeing.

This work presents significant upgrades to VAMPIRES, improving its capabilities as a high-contrast imager. These upgrades revolved around the deployment of two CMOS detectors with high-framerate photon-number-resolving capabilities, a suite of coronagraphs optimized for the visible, a new multiband imaging mode for simultaneous imaging in multiple filters, and a new achromatic liquid crystal polarization modulator. These upgrades significantly improve VAMPIRES' high-contrast capabilities, reaching deeper sensitivity limits with improved efficiency.

\section{The VAMPIRES instrument}\label{sec:design}

\begin{figure*}
    \centering
    \includegraphics[width=\textwidth]{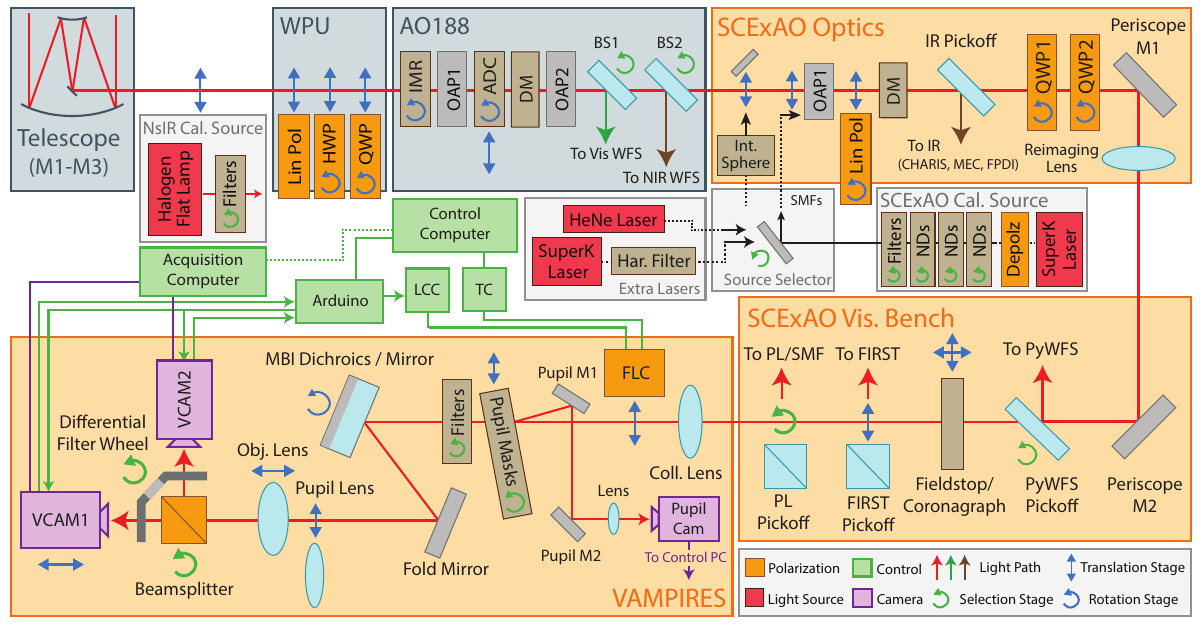}
    \caption{VAMPIRES Instrument Schematic including all common path instruments: the telescope, waveplate unit (WPU), AO188, and SCExAO. Some components are simplified or omitted for clarity and are not to scale. NsIR: Nasmyth infrared platform of the Subaru Telescope, HWP: half-wave plate, QWP: quarter-wave plate, IMR: image rotator (K-mirror), OAP: off-axis parabolic mirror, ADC: atmospheric dispersion corrector, DM: deformable mirror, BS: beamsplitter, PyWFS: pyramid wavefront sensor, PL: photonic lantern, SMF: single-mode fiber, FLC: ferroelectric liquid crystal, LCC: liquid-crystal controller, TC: temperature controller, ND: neutral density filter.\label{fig:schematic}}
\end{figure*}

The Visible Aperture-Masking Polarimetric Imager/Interferometer for Resolving Exoplanetary Signatures (VAMPIRES) is the visible-light (\SIrange{600}{775}{\nano\meter}) sub-instrument of the Subaru Coronagraphic Extreme Adaptive Optics testbed (SCExAO; \citealp{jovanovic_subaru_2015}). SCExAO is a platform for high-contrast imaging and technology development on the \SI{8.2}{\meter} Subaru telescope; the multi-stage adaptive optics achieves diffraction-limited imaging at visible wavelengths, and the large telescope diameter produces angular resolutions of $\sim$\SI{20}{\mas}. Observing modes and science use cases for VAMPIRES are detailed in \autoref{tbl:modes}.

\begin{deluxetable*}{p{1.2in}p{2.5in}p{3in}}
\tabletypesize{\small}
\tablecaption{VAMPIRES observing configurations.\label{tbl:modes}}
\tablehead{
    \colhead{Mode} & 
    \colhead{Description} &
    \colhead{Use cases}
}
\tablecolumns{3}
\startdata
\multicolumn{3}{l}{\textit{Filters}} \\
Standard & Open, 625-50, 675-50, 725-50, 750-50, 775-50 & Highest throughput with Open filter \\
Multiband & F610, F670, F720, F760 & Spectral differential imaging, color analysis, phase diversity \\
Narrowband & H$\alpha$, H$\alpha$-Cont, SII, SII-Cont & Narrowband spectral differential imaging, accreting protoplanets, stellar jets, stellar atmospheres \\
\hline \multicolumn{3}{l}{\textit{Polarimetry}} \\
Fast polarimetry & Triple-difference with HWP+FLC\newline (DIT$<$\SI{1}{\second}) & Circumstellar disks, sub-stellar companions, stellar atmospheres, stellar jets, solar system objects \\
Slow polarimetry & Double-difference with HWP & Circumstellar disks, stellar jets, solar system objects \\
\hline \multicolumn{3}{l}{\textit{Coronagraphy}} \\
No coronagraph & 5$\sigma$ contrast: $10^{\text{-}3}$ at \ang{;;0.1}, $10^{\text{-}4}$ at $>$\ang{;;0.4} & Stellar companions, circumstellar disks (with PDI), solar system objects \\
Classic Lyot & IWA: 37, 59, 105, or 150 \si{\mas};\newline 5$\sigma$ contrast: $10^{\text{-}4}$ at \ang{;;0.1}, $10^{\text{-}6}$ at $>$\ang{;;0.5} & Sub-stellar companions and circumstellar disks \\
Vector vortex & IWA: 61 \si{\mas}; 5$\sigma$ contrast: TBD\newline (not compatible with polarimetry) & Sub-stellar companions and circumstellar disks \\
\hline \multicolumn{3}{l}{\textit{Aperture masking}} \\
NRM interferometry & 7, 9, and 18-hole non-redundant masks, plus a redundant annulus mask (not compatible with coronagraphs) & Dust shells, stellar atmospheres, circumstellar disks \\
Redundant apodizing pupil & 8 to 35 $\lambda/D$ dark hole resilient to $<$\SI{1}{\radian} of low-wind effect (not compatible with coronagraphs) & Circumstellar disks and sub-stellar companions \\
\enddata
\end{deluxetable*}

\subsection{Telescope and Common Path Instruments}

SCExAO is mounted on one of the Nasmyth platforms of the Subaru telescope. The first common-path instrument is the facility waveplate unit (WPU; \citealp{watanabe_near-infrared_2018}). The WPU includes a linear polarizer, rotatable half-wave plate (HWP), and rotatable quarter-wave plate (QWP), which is used for polarimetric modulation and calibration (\autoref{sec:polarimetry}). Next is Subaru's facility AO instrument, which provides a first-level atmospheric turbulence correction. It includes a broadband atmospheric dispersion corrector (ADC), visible and near-infrared wavefront sensors (WFS), and a deformable mirror (DM). A \SI{600}{\nano\meter} longpass dichroic splits light between the WFS ($<$\SI{600}{\nano\meter}) and downstream instruments ($>$\SI{600}{\nano\meter}). The results presented in this paper were obtained with a 188-actuator DM used for this stage (AO188; \citealp{minowa_performance_2010}): it has since been replaced by a 3228-count actuator DM (AO3K; see \autoref{sec:futureprospects}). 

\subsection{SCExAO Common Path}

Light entering SCExAO is corrected with a 2000-actuator DM at up to \SI{3.6}{\kilo\hertz} speeds for extreme AO correction \citep{lozi_new_2020,ahn_scexao_2021}. SCExAO operates its deformable mirror independently from the facility AO in a cascading way, correcting the residual errors. The corrected beam is split with NIR light sent to a suite of sub-instruments, including CHARIS \citep{groff_charis_2015}, MEC \citep{steiger_probing_2022}, and FastPDI \citep{lozi_status_2020}, while visible light is reflected into a periscope and reimaging lens.

A portion of the light exiting the periscope is sent to the visible pyramid WFS (PyWFS; \citealp{lozi_visible_2019}). Various dichroic and gray beamsplitters are available, but the typical configuration uses an \SI{800}{\nano\meter} shortpass dichroic. The tilt angle of this dichroic ($\sim$\ang{20}) shifts the cut-off wavelength closer to $\sim$\SI{775}{\nano\meter}. Following this pickoff is a focal plane mount that hosts a \ang{;;3}x\ang{;;3} field stop and a suite of visible coronagraphs (\autoref{sec:coronagraphy}). Following the field stop are two gray beamsplitters for fiber-fed sub-instruments \citep{vievard_single-aperture_2023,vievard_photonic_2023}.

\subsection{VAMPIRES Optics}

The remaining light (\SIrange{600}{775}{\nano\meter}) is collimated into a \SI{7.06}{\milli\meter} diameter beam. This beam passes through a removable ferroelectric liquid crystal (FLC), a pupil mask wheel, and a filter wheel. The pupil wheel houses sparse aperture masks, neutral-density (ND) filters, and the coronagraphic Lyot stops. This wheel is tilted at $\sim$\ang{6} so that a separate pupil-viewing camera can image light reflected off the masks. The standard filter wheel houses five \SI{50}{\nano\meter}-wide bandpass filters for standard imaging and an empty slot for broadband observations. After the filter wheel is a baffle isolating the VAMPIRES fore-optics.

Next, the beam reaches the multiband imaging optics and the VAMPIRES objective lens. A pupil-imaging lens can be inserted with a flip mount. As the beam converges, it passes the beamsplitter wheel, which hosts a wire-grid polarizing cube and a gray non-polarizing cube. The beamsplitters can be removed for single-camera operation. The last device before the detectors is a differential filter wheel that swaps narrowband filters between each camera for narrowband spectral differential imaging \citep{uyama_high-contrast_2020}. The two VAMPIRES detectors are described in detail in \autoref{sec:detectors}.

\subsection{Multiband Imaging} \label{sec:mbi}

One of the main goals of the VAMPIRES upgrades was enabling spectral differential imaging (SDI) without adding dispersive optics. Dispersive optics, like in an integral-field unit (IFU), require space for a prism or grating, lenslets, and reimaging optics. Space is limited in VAMPIRES to only $\sim$\SI{20}{\centi\meter} after the pupil mask. We developed a technique for ``low-resolution dispersion'' using dichroic filters to form multiple images in a compact assembly called multiband imaging (MBI).

\begin{figure}
    \centering
    \includegraphics[width=\columnwidth]{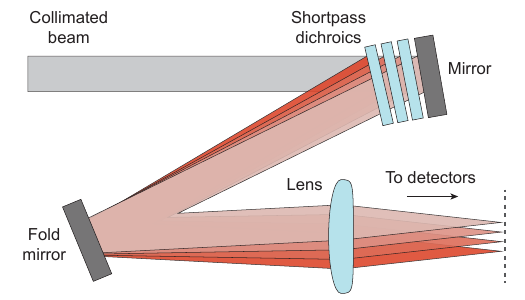}
    \caption{Schematic drawing of the multiband imaging principle. Broadband collimated light reaches the stack of tilted dichroics and mirrors. Light reflected by each dichroic creates a unique field in the focal plane due to the angle of incidence, transmitting the remaining bandpass onto the next dichroic. The final mirror reflects all remaining light through the stack.\label{fig:mbi_schematic}}
\end{figure}

The MBI technique is enabled by using multiple dichroics with unique angles-of-incidence (AOI) to split broadband collimated light into multiple fields imaged on the detector (\autoref{fig:mbi_schematic}). This approach takes advantage of the large sensor size of the detectors, which only uses 3\% of the detector for a \ang{;;3}x\ang{;;3} field of view (FOV)-- up to four FOVs can be tiled side-by-side before the objective lens starts to distort the off-axis images, which still only uses half the available detector area.

The optical design comprises three shortpass dichroic filters (\SI{750}{\nano\meter}, \SI{700}{\nano\meter}, and \SI{650}{\nano\meter}) plus a protected silver mirror to create four $\sim$\SI{50}{\nano\meter}-bandpass fields (\autoref{fig:filters}). The assembly can be rotated \ang{180} to become a fold mirror. A $\sim$\ang{10} protected silver fold mirror redirects the beam(s) towards the objective lens.

Ghost images are produced by repeated reflections between dichroic surfaces, which interfere with high-contrast observations. The geometry of the dichroic tilt angles was optimized so that all ghost images fall completely outside of the FOVs of the primary fields (\autoref{sec:ghosts}). Field stops are required to avoid overlap between fields, and all coronagraph masks include a field stop.

The required precision of the MBI dichroic tilt angles is on the order of $\sim$\ang{;;3} to ensure neighboring fields do not overlap. This corresponds to a linear precision of $\sim$\SI{1}{\micron} along the circumference of the \SI{25}{\milli\meter} diameter, which is $\sim$100 times finer than standard machining tolerances. Precision manufacturers NH Micro\footnote{\url{https://www.nhmicro.com/}} created titanium ring spacers and a custom lens tube using electrical-discharge machining, which has $\sim$\SI{1}{\micron} tolerances. When fully assembled, the spacers self-align in the lens tube, and interferometric analysis shows no optical distortions of the wavefront. 

\section{Detectors}\label{sec:detectors}

Fast, low-noise detectors are a key component for HCI, and a major component of the VAMPIRES upgrades revolved around improving the detectors. Previously, VAMPIRES used two Andor iXon Ultra 897 EMCCDs\footnote{\url{https://andor.oxinst.com/products/ixon-emccd-camera-series/ixon-ultra-897}}. Through electron multiplication, the effective root-mean-square (RMS) read noise can be lowered to $\sim$\SI{0.27}{\electron}, but the dynamic range is limited (\SI{64}{\decibel}). In addition, the Andor detectors have a maximum framerate of \SI{56}{\hertz}, which is not fast enough to freeze the atmospheric speckles, which evolve on timescales of $\sim$\SI{1}{\milli\second} above Maunakea \citep{kooten_climate_2022}. 

We upgraded VAMPIRES by replacing the two Andor EMCCDs with Hamamatsu ORCA-Quest qCMOS detectors.\footnote{\url{https://www.hamamatsu.com/us/en/product/cameras/qcmos-cameras/C15550-20UP.html}} The qCMOS detectors were chosen for their fast readout ($\sim$500 Hz at 536x536), low read noise ($\sim$ 0.25 - 0.4 $e^-$ RMS), and high dynamic range ($\sim$ 85 - 90 \si{\decibel}). The new detectors have two readout modes: ``fast'' and ``slow'' mode. In ``fast'' mode the RMS read noise is \SI{0.4}{\electron} and the allowed detector integration times (DIT) are \SI{7.2}{\micro\second} - \SI{1800}{\second} with a maximum framerate of \SI{506}{\hertz} for a \ang{;;3}x\ang{;;3} FOV crop. In ``slow'' mode the RMS read noise is $\sim$\SI{0.25}{\electron}, but the DIT is limited to $>$\SI{48}{\milli\second} (\SI{21}{\hertz}). The detectors have rolling-shutter readout, which causes the pixel rows to have slightly different acquisition times. This effect is exacerbated at the fastest exposure times using the electronic shutter. Detailed analysis of the rolling shutter effect is left for future work.

The imaging fore-optics were redesigned to accommodate the smaller detector pixel pitch (\SI{4.6}{\micron} now versus \SI{16}{\micron} previously), which made the F-ratio faster (F/21 now versus F/52 previously).

\subsection{Detector Characterization}

To determine the conversion between detector counts and electrons, we measured photon transfer curves using a calibration flat field source for both detectors in both readout modes. The fixed-pattern noise (FPN) was removed through frame-differencing, and then a linear model was fit to the average signal and the signal variance \citep{janesick_photon_2007,stefanov_cmos_2022}
\begin{equation}
    \sigma^2\left(\mu | k, \sigma_{RN}\right) = \frac{\mu}{k} + \sigma^2_{RN}.
\end{equation}
where $\mu$ is the average intensity in \si{adu}, $k$ is the gain in \si{\electron/adu}, and $\sigma_{RN}$ is the root-mean-square (RMS) read noise in \si{adu}.

The detector characteristics are reported in \autoref{tbl:detectors}. The readout noise per pixel was measured from the standard deviation of multiple bias frames and combined in quadrature to derive the RMS read noise. The detector photoresponse has non-linearities at low flux levels and around $2^{13}$ adu where the junction of 12-bit ADCs occurs \citep[see][]{strakhov_speckle_2023}. The non-linearities can be corrected to less than 1\% with an approximated function. The signal clips at $2^{16}$ adu, which is less than the full-well capacity of \SI{7000}{\electron} ($\sim$\SI{67,000}{adu}). The detector dark current is \SI{3.6e-3}{\electron/\second/\pixel} at sensor temperatures of \SI{-45}{\celsius} to \SI{-40}{\celsius}.

\subsection{Photon Number Resolution}

\begin{figure}
    \centering
    \script{pnr.py}
    \includegraphics[width=\columnwidth]{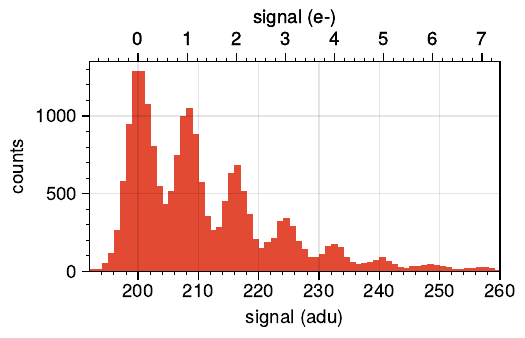}
    \caption{Histogram of $10^4$ pixels from a long exposure in ``slow'' mode ($\sigma_{RN}$=\SI{0.22}{\electron}). All signals are due to dark current. The Poisson photon number peaks are resolved.\label{fig:pch}}
\end{figure}

Since the qCMOS detectors have read noise below $\sim$\SI{0.3}{\electron}, they can resolve photon numbers, unambiguously detecting the number of individual photons hitting each pixel  \citep{starkey_determining_2016}. The photon number statistic is free from frame-to-frame noise (like readout noise) but requires accurate bias subtraction and flat-fielding for practical use. A derivation of the signal-to-noise ratio (S/N) gained by photon number counting is in \autoref{sec:pnr_derivation}. Photon number resolution is demonstrated in \autoref{fig:pch}, which shows a histogram of $10^4$ pixels from a long exposure dark frame. The photon peaks from Poisson statistics are resolved. The method from \citet{starkey_determining_2016} was used to determine the gain and quanta exposure directly from the histogram peaks, consistent with results from the photon transfer curves.

\subsection{qCMOS versus EMCCD}

\begin{figure}
    \centering
    \script{detector_snr.py}
    \includegraphics[width=\columnwidth]{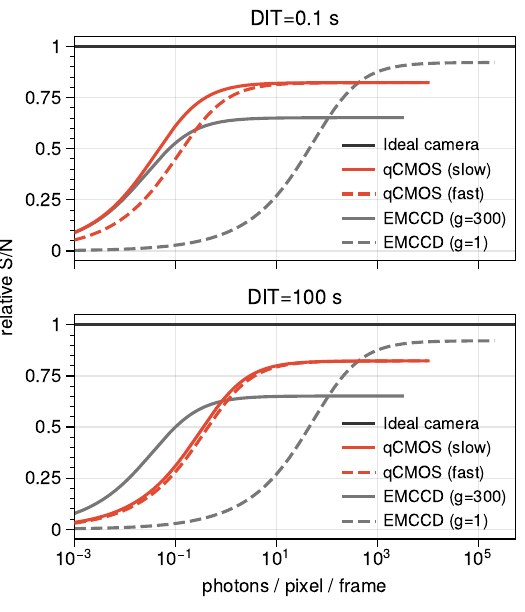}
    \caption{Theoretical normalized S/N curves for the new CMOS detectors (red curves) and the previous EMCCD detectors (gray curves). The curves are normalized to an ideal camera (only photon noise with perfect QE; black curve). The top plot shows a read noise-limited case, and the bottom plot shows a dark-limited case. Noise terms include read noise, dark noise, photon noise, and CIC plus excess noise factor for the EMCCDs.\label{fig:detector_snr_relative}}
\end{figure}

The relative sensitivity of the qCMOS detectors compared to the previous EMCCDs is shown in \autoref{fig:detector_snr_relative}. The number of detected photoelectrons is
\begin{equation}
    S_e = f \cdot t \cdot QE,
\end{equation}
where $f$ is the photon flux, $t$ is the detector integration time, and $QE$ is the bandpass-averaged quantum efficiency. Dark-current ($f_{DC}$) is included for both detectors, which is only a factor of exposure time--
\begin{equation}
    D_e = f_{DC} \cdot t.
\end{equation}

The S/N per frame for the qCMOS detectors contains shot noise, dark current, and read noise \citep{janesick_photon_2007,stefanov_cmos_2022}--
\begin{equation}
    S/N_\mathrm{CMOS} = \frac{S_e}{\sqrt{S_e + D_e + \sigma_{RN}^2}},
\end{equation}
where $\sigma_{RN}$ is the RMS read noise.

For the EMCCDs, extra noise terms unique to the electron-multiplication process \citep{harpsoe_bayesian_2012} are incorporated into
\begin{equation}
    S/N_\mathrm{EMCCD} = \frac{S_e}{\sqrt{\left(S_e + D_e + f_{CIC}\right) \cdot \gamma^2 + \left(\sigma_{RN}/g\right)^2}},
\end{equation}
where $g$ is the EM gain, $\gamma$ is the excess noise factor ($\sqrt{2}$ with EM gain), and $f_{CIC}$ is the clock-induced charge (CIC). CIC of $10^{-3}$ \si{\electron/\pixel/frame} was used based on empirical tests, and vendor values were used for everything else.

The qCMOS detectors have an order of magnitude worse dark current and slightly worse QE than the EMCCDs. Despite this, \autoref{fig:detector_snr_relative} shows that the qCMOS detectors perform better than the EMCCDs over a broad range of illuminations, especially at low-light levels. The qCMOS detectors are limited by the dark current shot noise at $\sim$\SI{15}{s} in ``slow'' mode, where the relative performance in the low-flux regime drops compared to the EMCCDs. In practice, though, the sky background is the limiting noise term at these exposure times, which has a larger effect on the S/N for the EMCCDs due to excess noise. Using the EMCCDs without electron multiplication is superior at high photon fluxes due to the lack of excess noise factor (compared to the EMCCD) combined with higher QE (compared to the qCMOS).

\begin{deluxetable}{lcccc}
\tabletypesize{\small}
\tablehead{
    \multirow{2}{*}{Parameter} &
    \multicolumn{2}{c}{qCMOS} & 
    \multicolumn{2}{c}{EMCCD} \\
    &
    \colhead{fast} &
    \colhead{slow} & 
    \colhead{g=1} &
    \colhead{g=300}
}
\tablecaption{VAMPIRES detector characteristics.\label{tbl:detectors}}
\startdata
Gain (\si{\electron/adu})  & 0.103  & 0.105  & 4.5  & 4.5  \\
RMS read noise (\si{\electron})      & 0.40   & 0.25   & 9.6 & 0.27 \\
Dark current (\si{\electron/\second/\pixel}) & \multicolumn{2}{c}{\num{3.6e-3}} & \multicolumn{2}{c}{\num{1.5e-4}} \\
Saturation limit$^a$ (\si{\electron}) & 6730 & 6860 & \num{1.5e5} & 980 \\
Dyn. range$^b$ (dB)      & 85     & 90     &  84  &  71    \\
Max framerate$^c$ (\si{\hertz}) & 506 & 21 & 56 & 56 \\
Pixel size (\si{\micron}) & \multicolumn{2}{c}{4.6} & \multicolumn{2}{c}{16} \\
Ave. QE$^d$ (\%) & \multicolumn{2}{c}{67} & \multicolumn{2}{c}{85} \\
\enddata
\tablecomments{(a) The qCMOS full well is larger than the bit-depth, so the saturation limit is the maximum input signal per pixel that can be recorded without clipping. (b) The input-referred dynamic range is derived from the saturation limit divided by the readout noise. (c) Maximum framerate for a 512$\times$512 pixel window. (d) Averaged over VAMPIRES bandpass (\SIrange{600}{800}{\nano\meter}).}
\end{deluxetable}


\subsection{Astrometric Calibration}

Previously, \citet{currie_images_2022} derived a plate scale of \SI{6.24\pm0.01}{\mas/\pixel} and parallactic angle offset of \ang{78.6\pm1.2} using a single epoch of observations of HD 1160B.  However, the changes in optics and detectors, in combination with the non-telecentricity of VAMPIRES (\autoref{sec:telecentricity}), prompted a new astrometric calibration for VAMPIRES.  We derive a new astrometric solution for each of the new detectors using a variety of binary star observations.

We observed six binary systems (Albireo, 21 Oph, HD 1160, HIP 3373, HD 137909, and HD 139341) with separations from \ang{;;0.12} to \ang{;;1.1} (\autoref{tbl:binaries}). The uncertainty in plate scale and position angle was dominated by the statistical uncertainty of the ephemerides rather than centroid precision. The observations of HD 1160 only used VCAM1 and were complicated by the process of PSF subtraction, which was required to detect the low-mass companion (see \autoref{sec:hd1160} for more details).

The derived instrument plate scale and parallactic angle offset are shown in \autoref{tbl:astrometry}. The instrument angle is defined as the pupil rotation of VAMPIRES to the image rotator zero point, which is calculated by
\begin{equation}
    \theta_\mathrm{inst} = \theta_\mathrm{off} + 180\text{\textdegree} - \theta_\mathrm{PAP},
\end{equation}
where $\theta_\mathrm{PAP}$ is the image rotator pupil offset of \ang{-39} (to align with the SCExAO entrance pupil). There is a \ang{180} change in position angle due to a parity flip compared to the previous astrometric solution for VAMPIRES \citep{currie_images_2022}.

Each camera was characterized separately with and without the multiband imaging dichroics (MBI), but there were no significant differences between the modes. Still, observers with the highest astrometric requirements should observe calibration targets during their run.

\begin{deluxetable}{lcccl}
\tabletypesize{\small}
\tablehead{
    \multirow{2}{*}{Object} &
    \colhead{$t_\mathrm{obs}$} &
    \colhead{$\rho$} &
    \colhead{$\theta$} &
    \multirow{2}{*}{Ref.} \vspace{-0.75em}\\
    &
    \colhead{(UT)} &
    \colhead{(\si{\mas})} &
    \colhead{(\textdegree)} &
}
\tablecaption{Visual binary ephemerides used for astrometric calibration.\label{tbl:binaries}}
\startdata
Albireo & \formatdate{27}{6}{2023} & \num{310\pm7.7} & \num{45.1\pm6.2} & [1] \\
21 Oph & \formatdate{8}{7}{2023} & \num{801\pm39} & \num{-53.0\pm3.1} & [2,3] \\
HD 1160 & \formatdate{11}{7}{2023} & \num{794.4\pm8.2} & \num{244.3\pm0.39} & [4] \\
HIP 3373 & \formatdate{30}{7}{2023} & \num{393\pm19} & \num{-0.7\pm3.1} & [5] \\
HD 137909 & \formatdate{30}{4}{2024} & \num{124.91\pm0.81} & \num{-166.0 \pm 0.23} & [6]\\
HD 139341 & \formatdate{30}{4}{2024} & \num{1131.9 \pm 3.0} & \num{178.15 \pm 0.24} &[7] \\
\enddata
\tablecomments{Separation ($\rho$) and position angle (East of North, $\theta$) are reported for the average observation time of the data.}
\tablerefs{[1]: \citet{drimmel_celestial_2021}, [2]: \citet{docobo_new_2007}, [3]: \citet{docobo_iau_2017}, [4]: \citet{bowler_population-level_2020} [5]: \citet{miles_iau_2017}, [6]: \citet{muterspaugh_phases_2010}, [7]: \citet{izmailov_orbits_2019}}
\end{deluxetable}

\begin{figure}
    \centering
    \script{astrometry_distributions.py}
    \includegraphics[width=\columnwidth]{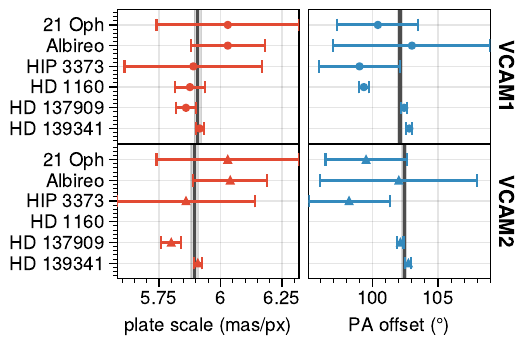}
    \caption{Results of astrometric characterization from each calibrator system shown as the mean with 1$\sigma$ error bars. The weighted mean and standard error of the weighted mean are shown with a gray vertical line and shaded contours. There are no VCAM2 data for HD1160.\label{fig:astrometry_results}}
\end{figure}

\begin{deluxetable}{lccc}
\tabletypesize{\small}
\tablehead{
    \multirow{2}{*}{Cam} &
    \colhead{pix. scale} &
    \colhead{$\theta_\mathrm{off}$} &
    \colhead{$\theta_\mathrm{inst}$} \vspace{-0.5em}\\
    &
    \colhead{(mas/px)} &
    \colhead{(\textdegree)} &
    \colhead{(\textdegree)}
}
\tablecaption{Astrometric characteristics of VAMPIRES.\label{tbl:astrometry}}
\startdata
VCAM1 & \num{5.908\pm0.014} & \num{102.10\pm0.15} & \num{-38.90\pm0.15} \\
VCAM2 & \num{5.895\pm0.015} & \num{102.42\pm0.17} & \num{-38.58\pm0.17} \\
\enddata
\tablecomments{The values and uncertainties are the variance-weighted mean and standard error from six observations (\autoref{fig:astrometry_results}).}
\end{deluxetable}

\section{Filters and Instrument Throughput}\label{sec:filters}

\subsection{Spectral Filters}
\begin{figure}
    \centering
    \script{filter_curves.py}
    \includegraphics[width=\columnwidth]{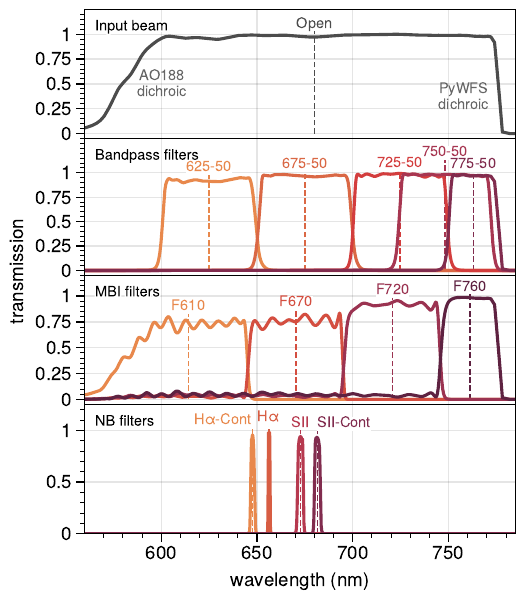}
    \caption{VAMPIRES filter transmission curves. All curves are normalized to the Open filter, the bandpass created from the dichroics upstream of VAMPIRES. The average wavelength for each filter is shown with a vertical dashed line.\label{fig:filters}}
\end{figure}

VAMPIRES has five \SI{50}{\nm} bandpass filters, four narrowband filters, an open broadband filter, and the dichroic filters in the multiband optic. All spectral filter transmission curves are compiled from manufacturer data and are shown in \autoref{fig:filters}. The filter characteristics and zero points are summarized in \autoref{tbl:filters}.

The Open filter is the widest and is constrained by the AO188 beamsplitter and the PyWFS pickoff. The multiband filters are defined by the consecutive transmissions/reflections through the stack of dichroics. The effective bandpasses are $\sim$\SI{50}{\nm} wide and have spectral leakage ($<$5\%) between some fields due to the imperfect dichroic transmissions. The narrowband filters come in pairs to enable differential imaging of the emission line and the local continuum. The H$\alpha$ filter is \SI{1}{\nm} wide and the H$\alpha$-continuum filter is \SI{2}{\nm} wide. The SII doublet and continuum filters are both \SI{3}{\nm} wide.

\begin{deluxetable*}{lcccccccccccc}
\centering
\tablehead{
    \multirow{2}{*}{Filter} &
    \colhead{$\lambda_\mathrm{ave}$} &
    \colhead{$\lambda_\mathrm{min}-\lambda_\mathrm{max}$} &
    \colhead{FWHM} &
    \colhead{$k$} &
    \colhead{Through.} &
    \colhead{$\mathrm{QE}_\mathrm{ave}$} &
    \multicolumn{3}{c}{Zero points} &
    \colhead{$C_{FD}$} \vspace{-0.5em}\\
    &
    \colhead{(\si{\nm})} &
    \colhead{(\si{\nm})} &
    \colhead{(\si{\nm})} &
    \colhead{(\si{mag/airmass})} &
    \colhead{(\%)} &
    \colhead{(\%)} &
    \colhead{(mag)} &
    \colhead{(\si{\electron/s})} &
    \colhead{(\si{Jy})} &
    \colhead{(\si{Jy.s/\electron})}
}
\tablecaption{VAMPIRES filter information.\label{tbl:filters}}
\startdata
Open & 680 & 580 - 776 & 196 & 0.06 & 6.1 & 68 & 25.8 & \num{2.2e+10} & 3550 & \num{1.6e-07} \\
625-50 & 625 & 601 - 649 & 49 & 0.08 & 5.1 & 74 & 24.3 & \num{5.3e+09} & 3440 & \num{6.6e-07} \\
675-50 & 675 & 651 - 699 & 49 & 0.05 & 8.6 & 67 & 24.6 & \num{7.0e+09} & 2990 & \num{4.3e-07} \\
725-50 & 725 & 700 - 749 & 49 & 0.04 & 11.7 & 61 & 24.7 & \num{7.9e+09} & 2800 & \num{3.6e-07} \\
750-50 & 748 & 724 - 776 & 48 & 0.03 & 13.0 & 58 & 24.7 & \num{7.7e+09} & 2680 & \num{3.6e-07} \\
775-50 & 763 & 750 - 776 & 26 & 0.03 & 7.3 & 57 & 23.2 & \num{2.1e+09} & 2610 & \num{1.6e-06} \\
\cutinhead{Multiband}
F610 & 614 & 580 - 645 & 65 & 0.09 & 3.0 & 77 & 24.0 & \num{3.9e+09} & 4810 & \num{1.2e-06} \\
F670 & 670 & 646 - 695 & 49 & 0.06 & 7.4 & 68 & 24.3 & \num{5.2e+09} & 3150 & \num{6.1e-07} \\
F720 & 721 & 696 - 745 & 50 & 0.04 & 9.2 & 62 & 24.6 & \num{6.7e+09} & 2920 & \num{4.4e-07} \\
F760 & 761 & 746 - 776 & 30 & 0.03 & 6.0 & 57 & 23.6 & \num{2.7e+09} & 2830 & \num{1.0e-06} \\
\cutinhead{Narrowband}
H$\alpha$ & 656.3 & 655.9 - 656.7 & 0.8 & 0.06 & 5.5 & 69 & 19.6 & \num{7.2e+07} & 2170 & \num{3.3e-05} \\
H$\alpha$-Cont & 647.7 & 646.7 - 648.6 & 1.9 & 0.07 & 6.0 & 70 & 20.9 & \num{2.3e+08} & 3210 & \num{1.4e-05} \\
SII & 672.7 & 671.2 - 674.1 & 2.9 & 0.05 & 10.8 & 67 & 21.8 & \num{5.5e+08} & 3090 & \num{5.7e-06} \\
SII-Cont & 681.5 & 680.0 - 683.0 & 3.0 & 0.05 & 8.9 & 66 & 21.6 & \num{4.3e+08} & 3040 & \num{7.0e-06}
\enddata
\tablecomments{$\lambda_\mathrm{ave}$: average wavelength, $\lambda_\mathrm{min}$: minimum wavelength with at least 50\% throughput, $\lambda_\mathrm{max}$: maximum wavelength with at least 50\% throughput, FWHM: full width at half-maximum, $k$: average atmospheric extinction from \citet{buton_atmospheric_2013}, Through: estimated instrument throughput (includes: telescope, excludes: beamsplitter, FLC, atmosphere), $\mathrm{QE}_\mathrm{ave}$: average detector quantum efficiency, $C_{FD}$: flux density conversion factor. Zero points are in Vega magnitudes and map to instrument flux without a beamsplitter in \si{\electron/s}. Zero points, instrument throughput, and conversion factors were estimated from spectrophotometric standard stars. All values assume there is no beamsplitter.}
\end{deluxetable*}

\subsection{Neutral Density Filters}

Two broadband reflective neutral density filters in the pupil mask wheel of VAMPIRES can be used to avoid saturating bright stars. Because these masks are in the pupil wheel, they cannot be used simultaneously with sparse aperture masks or Lyot stops. The optical density was measured for both filters: the ND10 has an optical density of \num{1.0}, and the ND25 has an optical density of \num{2.33}. The polarimetric effects of these filters have not been characterized and they are not recommended for use during polarimetric observations.

\subsection{Beamsplitter Throughput}

VAMPIRES has two beamsplitters: a wire-grid polarizing beamsplitter cube (PBS) and a non-polarizing beamsplitter cube (NPBS). The beamsplitter throughput was measured in the Open filter and compared to the flux without a beamsplitter (\autoref{tbl:bs}). To avoid polarization biases, a linear polarizer was inserted after the internal source and rotated in steps of \ang{5} from \ang{0} to \ang{180}, taking the average value from this range as the throughput.

\begin{deluxetable}{lccc}
\tablehead{
    \multirow{2}{*}{Name} &
    \multicolumn{3}{c}{Throughput (\%)} \\
    &
    \colhead{VCAM1} &
    \colhead{VCAM2} &
    \colhead{Sum}
}
\tablecaption{VAMPIRES beamsplitter throughput measurements.\label{tbl:bs}}
\startdata
Open & \num{100} & - & \num{100} \\
PBS & \num{36.76+-0.04} & \num{42.88+-0.04} & \num{79.64+-0.06} \\
NPBS & \num{43.95+-0.04} & \num{45.35+-0.04} & \num{89.31+-0.07} \\
\enddata
\tablecomments{PBS: Polarizing beamsplitter, NPBS: Non-polarizing beamsplitter.}
\end{deluxetable}

\subsection{Filter Color Corrections}

The filters in VAMPIRES do not follow any standard photometric system, although the Open filter is similar in central wavelength and width to the Johnson-Cousins R band \citep{bessell_ubvri_1979}. When using photometric values from online catalogs, the magnitudes must be converted from the source filter into the respective VAMPIRES filter, which depends on the spectrum of the observed star.

\begin{figure}
    \centering
    \script{color_correction.py}
    \includegraphics[width=\columnwidth]{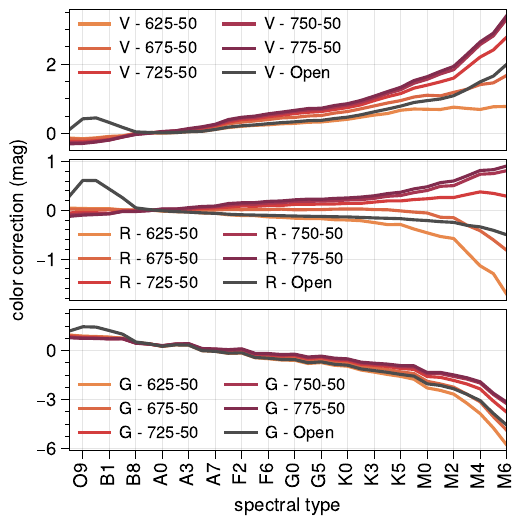}
    \caption{Color correction values for selected filters into the VAMPIRES bandpass filters. All correction values are tabulated from main-sequence Pickles stellar models. The spectral types are labeled on the x-axis. (Top) Johnson-Cousins V (Middle) Johnson-Cousins R (Bottom) Gaia G.\label{fig:color_correction}}
\end{figure}

The Pickles stellar models \citep{pickles_stellar_1998} were used to measure the color correction for varying spectral types of main-sequence stars. For each model spectrum, \texttt{pysynphot} was used to calculate the color correction in Vega magnitude from Johnson-Cousins V, R, and Gaia G to the respective VAMPIRES filter (\autoref{fig:color_correction}). The Johnson V filter deviates up to \num{1} magnitude fainter than the VAMPIRES filters for later spectral types, the R filter has consistent colors compared to VAMPIRES except for the coolest stars, and the G filter can deviate up to \num{5} magnitudes brighter compared to VAMPIRES, depending on the spectral type.

\subsection{Absolute Flux Calibration}

During commissioning, spectrophotometric standard stars HR 718 ($m_{\rm V}=$ \SI{4.273}) and HD 19445 ($m_{\rm V}=$ \SI{8.096}) were observed in each filter to calibrate the photometric zero points in \si{Jy} and \si{\electron/\second} \citep{hamuy_southern_1992,hamuy_southern_1994,stone_spectrophotometry_1996,zacharias_fourth_2013}.

The absolute flux calibration is summarized by the $C_{FD}$ coefficient, following \citet{gordon_james_2022}. This flux density conversion factor converts between instrument flux (in \si{\electron/\second}) and \si{Jy}. The instrument flux was measured using aperture photometry and corrected for atmospheric extinction using an empirical model from  \citet{buton_atmospheric_2013}. The conversion factors are derived from synthetic photometry of the calibrated stellar spectra.

The photometric zero points are mapped from Vega magnitudes to instrument flux in \si{\electron/\second} without a beamsplitter. The traditional zero point can be derived by incorporating each detector's beamsplitter throughput and camera gain rather than specifying all \num{280} unique combinations.

The zero points are used by taking the aperture flux in \si{\electron\per\second}, corrected for any beamsplitter or FLC throughputs, and converting to magnitudes with the zero point magnitude:
\begin{equation}
    m_{\rm filt}=\mathrm{ZP}_\mathrm{filt} - 2.5\log{\left(f\right)},
\end{equation}
or, equivalently, with the zero point flux
\begin{equation}
    m_{\rm filt}=-2.5\log{\left(f/f_\mathrm{ZP}\right)}.
\end{equation}
The surface brightness, $\Sigma_{\rm filt}$, can be calculated by dividing the calibrated flux by the area of a pixel ($\Omega_{px}=$\SI{3.5e-5}{arcsec^2/\pixel}). For Vega magnitudes, instead, use
\begin{equation}
    \label{eqn:surf_bright}
    \Sigma_{\rm filt} = m_{\rm filt} + 2.5\log{\left(\Omega_{px}\right)} = m_{\rm filt}- 11.1~\mathrm{mag/arcsec}^2.
\end{equation}

The accuracy of the zero points and conversion factors is constrained by atmospheric transmission variations along with dust accumulation and coating degradation of the telescope mirrors. Imaging a spectrophotometric standard at multiple elevation angles for observations demanding the highest photometric precision will give better accuracy.

\subsection{Limiting Magnitudes}

Under excellent conditions, the detector integration time and saturation limit set the brighter end of the magnitude range. The fastest detector integration time is \SI{7.2}{\micro\second} in ``fast'' mode and \SI{48}{\milli\second} in ``slow'' mode (for the standard imaging mode with a \ang{;;3}x\ang{;;3} FOV). For these calculations, the peak count limit is set to \SI{60000}{adu}, and it is assumed that the brightest pixel in the PSF will be 30\% the full aperture flux. The highest-throughput scenario (``Open'' filter, no beamsplitter, no FLC; \autoref{tbl:filters}) translates the limiting instrumental flux into $m_{\rm Open}$ = 2.1 in ``fast'' mode and $m_{\rm Open}$ = 11.7 in ``slow'' mode. The AO wavefront sensors constrain the faint observing limits: the current guider limits of AO188 are $m_{\rm R}$ $<$ 16, and the wavefront sensor limits for diffraction-limited imaging at visible wavelengths with SCExAO are roughly $m_{\rm R}$ $<$ 10 \citep{ahn_development_2023}.

The lunar phase primarily determines the sky background above Maunakea in the visible. The peak surface brightness is $\Sigma_{\rm R}=$\SI{17.9}{mag/arcsec^2} \citep{roth_measurements_2016}. Assuming, again, the highest-throughput observing configuration and using a pixel size of \SI{5.9}{mas}, this corresponds to an instrument flux of \SI{5e-2}{\electron/\pixel/\second}. This is an order of magnitude higher than the dark signal (\autoref{tbl:detectors}) and is the limiting noise term for exposures longer than \SI{1.2}{\second}.

\section{The VAMPIRES point-spread function}\label{sec:psf}

One of the key parameters of a high-contrast imaging instrument is the instrumental point-spread function (PSF) and how close the on-sky performance reaches the ideal PSF. The observed VAMPIRES PSF quality varies depending on the residual wavefront error and the exposure time. Because the PSF fluctuates rapidly from residual atmospheric wavefront errors, integrating longer than the speckle coherence time ($\sim$\SI{4}{\milli\second}, \citealp{kooten_climate_2022}) blurs the PSF. The speckled diffraction pattern can be ``frozen'' with fast exposure times, enabling lucky imaging. The high-speed detectors in VAMPIRES can easily take exposures faster than the coherence timescale, which makes it well-suited for this technique.

\begin{figure}
    \centering
    \script{onsky_psfs.py}
    \includegraphics[width=\columnwidth]{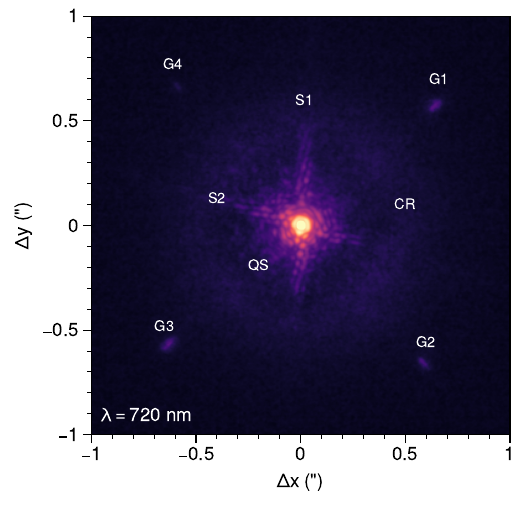}
    \caption{On-sky images of \textit{HD 191195} in the F720 filter (DIT=\SI{2.3}{\milli\second}). Data has been aligned and collapsed and is shown on a log scale with clipping to emphasize the fainter regions of the image. (G1-G4) passive speckles created by the diffractive gridding of the SCExAO deformable mirror. (S1-S2) Spider diffraction spikes. (CR) the SCExAO control radius, where a speckle halo creates an apparent dark hole. (QS) quasi-static speckles created by residual and non-common path wavefront errors.\label{fig:onsky_psf}}
\end{figure}

An annotated PSF in the F720 filter of HD 191195 after post-processing is shown in \autoref{fig:onsky_psf}. This data has a FWHM of \SI{20}{\mas} and a Strehl ratio\footnote{The Strehl ratio is measured using a broadband synthetic PSF normalized to the photometric flux in a \ang{;;0.5} radius aperture. The peak is subsampled with a Fourier transform, as in method seven (7) of \citet{roberts_is_2004}.} of $\sim$52\%, which is typical performance for SCExAO in average conditions (seeing of \ang{;;0.8}).

\subsection{PSF Features}
The PSF image in \autoref{fig:onsky_psf} shows typical features of high S/N VAMPIRES observations. The center of the frame contains the central core and the first few Airy rings, which are slightly asymmetric due to wavefront errors. Beyond this are the quasi-static speckles (QS), which slowly evolve due to non-common path errors \citep{soummer_speckle_2007,ndiaye_calibration_2013}.

The PSF contains two large, bright diffraction spikes from the secondary mirror support structure (spiders; S1-S2). These spikes reduce image contrast if they align with a target in the field, but using coronagraphy (\autoref{sec:coronagraphy}) or exploiting field rotation over time can mitigate the effects of these spikes.

The SCExAO deformable mirror has 45 actuators across the pupil, which limits the highest spatial frequency corrected by the AO loop to $\sim$22.5$\lambda/D$. Typically, this is referred to as the control radius or ``dark hole,'' visible in the PSF image (CR). The dark hole is closer to a square in this image, determined by the modal basis used for wavefront sensing and control. The DM also acts like a diffractive element due to the gridding of the surface, which creates copies of the stellar PSF at four locations orthogonal to the orientation of the DM (G1-G4).

\subsection{Seeing and the Low-Wind Effect}
\begin{figure}
    \centering
    \script{bad_psfs.py}
    \includegraphics[width=\columnwidth]{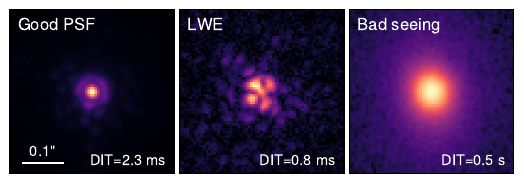}
    \caption{(Left) High-framerate image in good conditions shows a high-quality PSF (Strehl $\sim$52\%). (Middle) high-framerate image highlighting LWE. (Right) a longer exposure image in poor seeing conditions is blurred into a Moffat-style profile. All images are shown with a square-root stretch with separate limits.\label{fig:bad_psfs}}
\end{figure}

The low-wind effect (LWE) is caused by piston wavefront discontinuities across the telescope pupil spiders  \citep{milli_low_2018}. The effect in the focal plane is ``splitting'' of the PSF (\autoref{fig:bad_psfs}). Current wavefront sensors struggle to detect LWE, which periodically occurs in almost every VAMPIRES observation, although it can be mitigated to an extent through frame selection. The other PSF image is of a fainter star ($m_{\rm R}=$8.1) with \SI{0.5}{\second} exposures. The PSF is fuzzy with a FWHM of \SI{75}{\mas} and low S/N resulting from the blurring of the residual speckle pattern throughout the integration.

\subsection{Lucky Imaging}

\begin{figure*}
    \centering
    \script{lucky_imaging.py}
    \includegraphics[width=\textwidth]{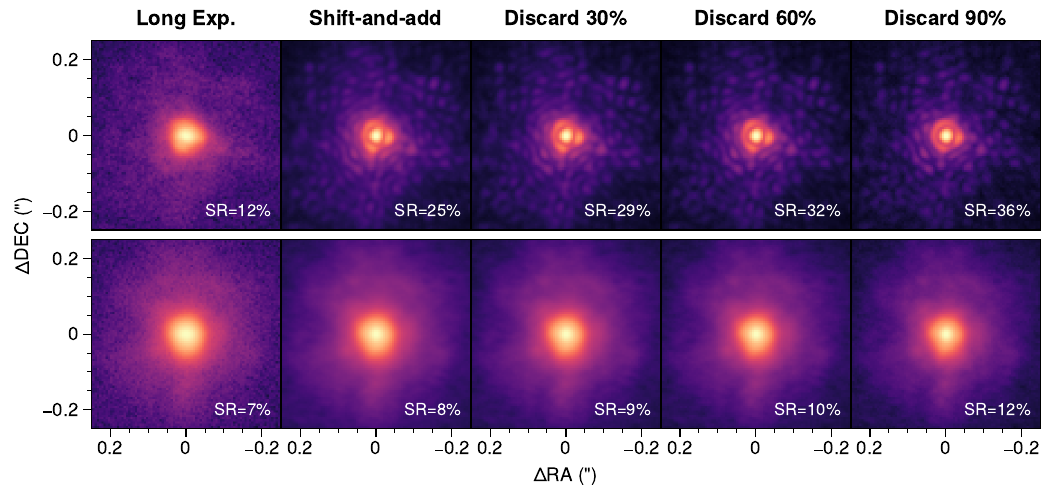}
    \caption{Post-processing data with lucky imaging. Each frame is shown with a log stretch and separate limits. (Top) high-framerate (\SI{500}{\hertz}) data in median seeing conditions ($m_{\rm R}=4.3$). (Bottom) low-framerate (\SI{2}{\hertz}) data in mediocre seeing conditions ($m_{\rm V}=9.2$). (Long Exp.) a mean combination without alignment, simulating a long exposure. (Shift-and-add) co-registering each frame before collapsing. (Discarding X\%) same as shift-and-add but discarding a percentage of data based on the Strehl ratio.\label{fig:lucky_imaging}}
\end{figure*}

\autoref{fig:lucky_imaging} shows two observing scenarios: high framerate data (\SI{500}{\hertz}) in median seeing conditions and low framerate data (\SI{2}{\hertz}) in poor seeing conditions. A long exposure was simulated by taking the mean image; then, the data was registered using centroids. Progressively, from left to right, more frames are discarded based on the Strehl ratio, which shows that frame selection increases the PSF quality, but in poor conditions, it lowers the overall S/N for little gain.

\subsection{Radial Profiles}
\begin{figure*}
    \centering
    \script{psf_profiles.py}
    \includegraphics[width=\textwidth]{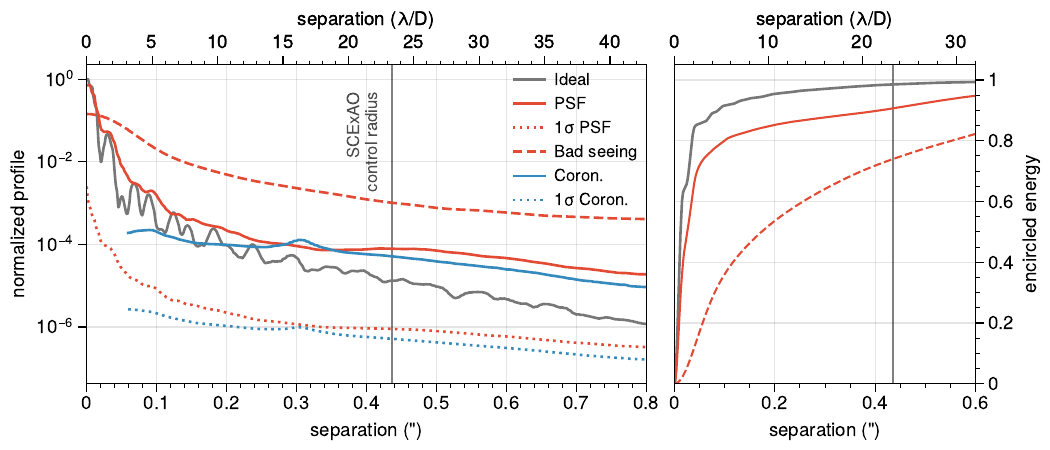}
    \caption{Radial profiles and encircled energy in different observing scenarios. (Left) radial profiles normalized to an ideal PSF (gray), a high-quality on-sky PSF (the same as \autoref{fig:onsky_psf}, solid red), a poor-quality on-sky PSF (the same as \autoref{fig:bad_psfs}, dashed red), and an on-sky coronagraphic PSF with the \SI{59}{\mas} IWA (solid blue). The $1\sigma$ raw contrast curves are shown for the good PSF (dotted red) and the coronagraphic PSF (dotted blue). The control radius of the SCExAO DM is marked with a vertical gray line. (Right) the encircled energy normalized to an ideal PSF with a max radius of \ang{;;1} (gray), a good on-sky PSF (solid red), and a poor-quality PSF (dashed red). The control radius of the SCExAO DM is marked with a vertical gray line.\label{fig:onsky_psf_profiles}}
\end{figure*}

The radial profiles of a high-quality PSF, a seeing-limited PSF, and a synthetic instrument PSF are shown in \autoref{fig:onsky_psf_profiles}. The ideal PSF is normalized to a peak value of 1 and the other profiles are normalized to the ideal PSF using a \ang{;;1} aperture sum. The cumulative aperture sums, or encircled energy, are normalized to a value of 1 at \ang{;;1} and shown alongside the radial profiles. The good, on-sky PSF closely matches the profile of the theoretical PSF until about 15$\lambda/D$. The nulls of the radial pattern are not as deep as the model due to quasi-static speckles. Past 15$\lambda/D$, the control radius of the SCExAO DM is clear due to the residual atmospheric speckle halo.

\section{Coronagraphy}\label{sec:coronagraphy}

Coronagraphy is another key tool for HCI because it reduces the stellar diffraction pattern and attenuates the photon noise from the on-axis PSF. This is essential for observing faint objects near bright stars, such as exoplanets or circumstellar disks, by significantly reducing the glare from the star itself. Four visible-light Lyot-style coronagraphs were installed in VAMPIRES \citep{lucas_visible-light_2022}. These coronagraphs provide up to three orders of magnitude of attenuation of the stellar PSF, which, when combined with extreme AO correction and low-noise detectors, enables contrasts $<10^{-6}$ (\autoref{fig:onsky_psf_profiles}).

\subsection{Focal Plane Masks}

The four Lyot-style focal plane masks are partially transmissive (0.6\%) circular dots with inner working angles (IWA) of two, three, five, and seven resolution elements ($\lambda/D$). The focal plane masks are mounted in a three-axis translation stage for fine alignment. In addition, a double-grating vector vortex coronagraph (DGVVC; \citealp{doelman_falco_2023}) was deployed in November 2023. This mask is currently being evaluated.

\begin{figure}
    \centering
    \script{iwa.py}
    \includegraphics{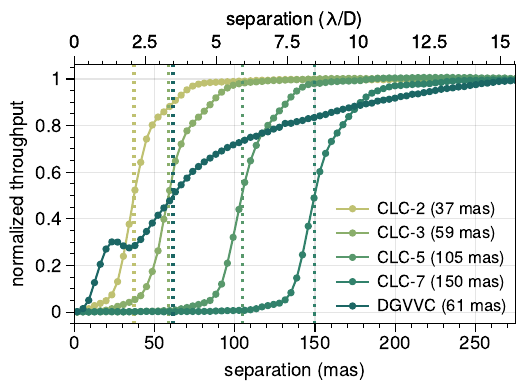}
    \caption{Normalized off-axis throughput for the VAMPIRES coronagraph masks. The throughputs are normalized to a range between 0 and 1. The inner working angle (IWA) is marked with vertical dotted lines at the point where the throughput reaches 50\%. $\lambda/D$ is determined from the average wavelength of the bandpass (\SI{680}{\nano\meter}).\label{fig:iwa}}
\end{figure}

The mask IWAs were measured by rastering the SCExAO internal calibration source across the focal plane mask and measuring the photometric throughput. The normalized throughput curves and IWA are shown in \autoref{fig:iwa}, and the results are summarized in \autoref{tbl:coronagraph}.

In practice, the CLC-2 mask is too small for on-sky observations-- the residual tip-tilt in average conditions causes the PSF to leak out of the side of the mask. Due to the large size of the CLC-7 mask and because the Lyot stop was tuned to the CLC-3 and CLC-5 masks, the CLC-7 mask produces some diffraction that is not rejected by the Lyot stop. In poor observing conditions, this can create speckles that appear pinned to the edge of the mask. The CLC-3 mask is well-suited for good conditions (seeing $<$\ang{;;0.6}) or for polarimetric observations (thanks to the efficiency of PDI). The CLC-5 mask offers a good balance between robust diffraction control and IWA for all other scenarios.

\begin{deluxetable}{lcc}
\tablehead{\colhead{Name} & \colhead{Radius (\si{\micron})} & \colhead{IWA (\si{\mas})}}
\tablecaption{VAMPIRES coronagraph mask specifications.\label{tbl:coronagraph}}
\startdata
CLC-2 & 46 & 37 \\
CLC-3 & 69 & 59 \\
CLC-5 & 116 & 105 \\
CLC-7 & 148 & 150 \\
DGVVC & 7 & 61 \\
\enddata
\tablecomments{The radius of the DGVVC is the radius of the central defect amplitude mask, which is resolved.}
\end{deluxetable}

\subsection{Pupil Masks}

\begin{figure}
    \centering
    \script{pupil_masks.py}
    \includegraphics{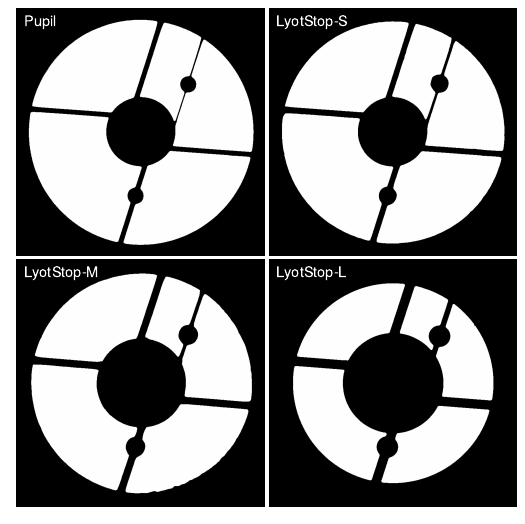}
    \caption{Images of the VAMPIRES pupil and coronagraphic pupil masks illuminated by a flat lamp. Images have been binarized to sharpen the edges. Each mask is labeled in the top left. The pupil image displays the SCExAO pupil, which includes masks for broken deformable mirror actuators.\label{fig:pupil_masks}}
\end{figure}

VAMPIRES has three Lyot stop masks to reject the light diffracted by the coronagraphic focal plane mask. The first Lyot stop is described in \citet{lucas_visible-light_2022}. This mask is coated with gold for high reflectivity, aiding alignment with the pupil camera and for a future low-order wavefront sensor. Two new masks were designed with higher throughput and deployed in January 2024 (PI: Walk, A.). These masks were laser-cut from \SI{1}{mm}-thick aluminum foil sheets, which are not smooth or reflective enough to use with the pupil-viewing camera. \autoref{fig:pupil_masks} shows all three Lyot stop masks imaged with a flat lamp alongside the VAMPIRES pupil. The mask specifications and throughput measurements are listed in \autoref{tbl:lyot_stops}.

\begin{deluxetable}{lcccc}
\tabletypesize{\small}
\tablehead{\multirow{2}{*}{Name} & \colhead{$D_\mathrm{in}$} & \colhead{$D_\mathrm{out}$} &  \colhead{$T_{geom}$} & \colhead{$T_{phot}$}  \vspace{-0.75em}\\
& \colhead{(\si{\milli\meter})} & \colhead{(\si{\milli\meter})} & \colhead{(\%)} & \colhead{(\%)}
}
\tablecaption{Instrument pupil and  Lyot stop specifications.\label{tbl:lyot_stops}}
\startdata
Telescope Pupil & 2300 & 7950 & - & -\\
SCExAO Pupil & 5.27 & 17.92 & - & - \\
VAMPIRES Pupil & 2.08 & 7.06 & 100 & 100 \\
\tableline
LyotStop-S & 2.14 & 7.06 & 95.9 & 89.2 \\
LyotStop-M & 2.82 & 6.99 & 85.7 & 79.7 \\
LyotStop-L & 3.16 & 6.33 &  61.5 & 57.2 \\
\enddata
\tablecomments{The telescope pupil is the effective pupil with the IR M2 (which undersizes the \SI{8.2}{\meter} aperture).}
\end{deluxetable}

\subsection{Redundant Apodizing Pupil Mask}

\begin{figure}
    \centering
    \script{rap.py}
    \includegraphics{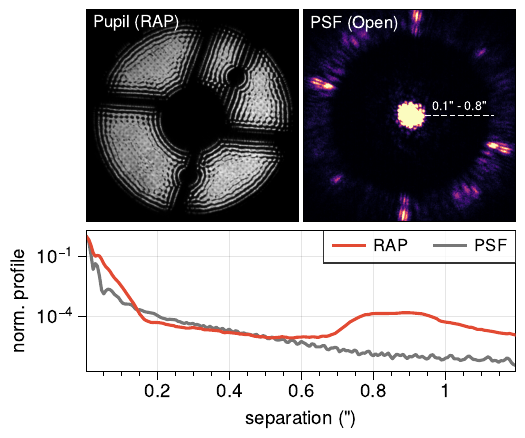}
    \caption{The redundant apodized pupil (RAP) mask. (Top left) a pupil image of the mask shows the apodization pattern. (Top right) a \ang{;;2}x\ang{;;2} FOV of the PSF produced by the RAP using the SCExAO internal source in the Open filter. The image is stretched to emphasize the dark hole from approximately \ang{;;0.1} to \ang{;;0.8}. (Bottom) a normalized radial profile of the top right PSF in log scale.\label{fig:rap}}
\end{figure}

A redundant apodized pupil mask (RAP; \citealp{leboulleux_redundant_2022,leboulleux_coronagraphy_2024}) was deployed on VAMPIRES in September 2023. The mask shapes the pupil and was designed to create a deep $10^{\text{-}6}$ contrast dark zone in an annulus from 8 to 35$\lambda/D$. The apodization pattern is resilient to up to  $\sim$\SI{1}{rad} of LWE, where the coherence of the PSF core starts to break down. The mask is designed with a trade-off between the size of the dark zone, the LWE resilience, and the mask's overall throughput, which is $\sim$25\%. The RAP mask, PSF, and radial profile are shown in \autoref{fig:rap}. Further testing and verification of this mask is future work.

\subsection{Coronagraphic Point-Spread Function}

\begin{figure}
    \centering
    \includegraphics[width=\columnwidth]{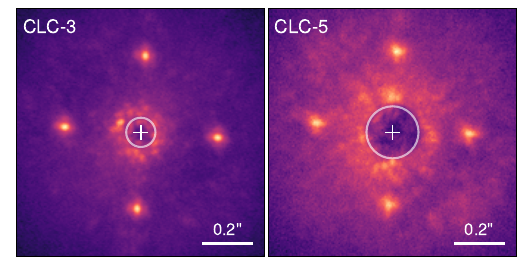}
    \caption{On-sky coronagraphic PSFs using the multiband imaging mode. (left) The CLC-3 mask from polarimetric observations of HD 163296. (right) the CLC-5 mask from polarimetric observations of HD 169142, which had slightly worse seeing conditions. Both are aligned and coadded from the F720 frame of single data cube, cropped to the inner \ang{;;1} FOV, and shown with a logarithmic stretch and separate limits for each image. The calibration speckles are $\sim$15.5$\lambda/D$ from the center. The coronagraph mask is overlaid with a white circle, and the approximate location of the star is marked with a cross.\label{fig:coro_psfs}}
\end{figure}

\autoref{fig:coro_psfs} shows aligned and coadded on-sky coronagraphic PSFs for the CLC-3 and CLC-5 masks. The PSFs show some common features: a dark hole in the DM control region, calibration speckles, and a residual speckle halo slightly blurred from the coadding. The CLC-3 image shows a secondary diffraction ring past the edge of the mask. The control region in the CLC-5 image has worse contrast due to poorer atmospheric conditions that night. The radial profile for the CLC-3 mask is shown in blue alongside the non-coronagraphic PSFs in \autoref{fig:onsky_psf_profiles}.

\subsection{Calibration Speckles}\label{sec:astrogrid}

VAMPIRES uses the SCExAO deformable mirror to create calibration speckles (``astrogrid'') for precise astrometry and photometry of the star behind the coronagraph mask \citep{sahoo_precision_2020}. The calibration speckles produce an ``X'' with theoretical separations of 10.3, 15.5, or 31.0 $\lambda/D$. The relative brightness of these speckles to the on-axis PSF allows for the precise photometry of coronagraphic images.

In the linear wavefront error regime ($<$\SI{1}{\radian}), the brightness of a speckle is a quadratic function of the optical path difference (OPD) \citep{jovanovic_artificial_2015,currie_laboratory_2018,chen_post-processing_2023} and, therefore, the ratio of the satellite spot flux to the central star (contrast) is
\begin{equation}
    \label{eqn:astrogrid}
    \frac{f_p}{f_*}\left( \lambda, A_{DM}\right) \propto \left(A_{DM} / \lambda\right)^2,
\end{equation}
where $A_{DM}$ is the mechanical displacement of the DM, $\lambda$ is the wavelength of light. 

\begin{figure}
    \centering
    \script{astrogrid_photometry.py}
    \includegraphics[width=\columnwidth]{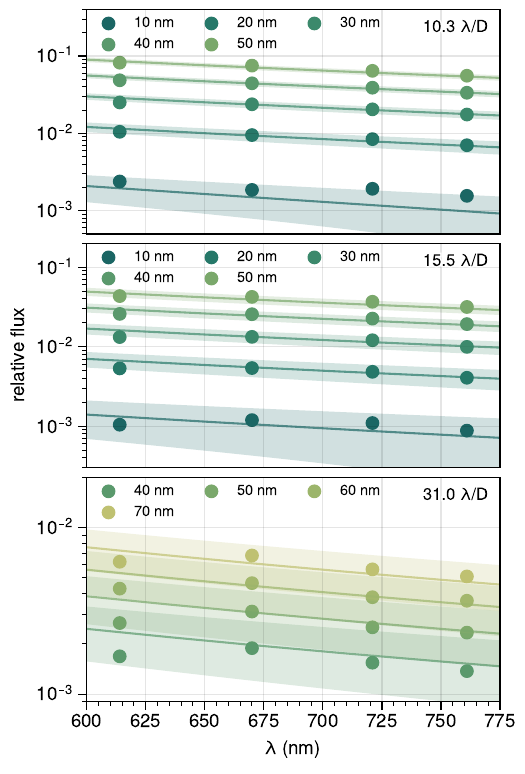}
    \caption{Relative photometric flux of the astrogrid calibration speckles. Each plot corresponds to a speckle pattern, and each curve corresponds to a given pattern amplitude (in mechanical microns applied to the DM) The best-fitting models with 1-$\sigma$ uncertainty bands are shown for each configuration. \label{fig:astrogrid_photometry}}
\end{figure}

The scaling coefficients for the most commonly used astrogrid patterns (10.3, 15.5, and 31 $\lambda/D$) were fit using multiband observations of the internal calibration source. When trying to fit \autoref{eqn:astrogrid}, a better agreement was found using a model containing an additional linear OPD term--
\begin{equation}
    \label{eqn:astrogridmod}
    \frac{f_p}{f_*}\left( \lambda, A_{DM} | c_0, c_1 \right) = c_0 \cdot \left(A_{DM} / \lambda\right) + c_1 \cdot \left(A_{DM} / \lambda\right)^2,
\end{equation}
which was fit using weighted linear least-squares, shown in \autoref{tbl:astrogrid} and \autoref{fig:astrogrid_photometry}. The 31$\lambda/D$ separation pattern has very low relative flux for the pattern amplitude due to the influence function of the DM, which has a weaker response for high spatial frequencies.

A trade-off exists between photometric and astrometric precision of the calibration speckles with increased contrast due to photon noise around them. The amplitude and separation of the astrogrid pattern should be tuned for optimal contrast over most of the control region. The wings of the PSF quickly fall below contrasts of $\sim10^{\text -2}$ and the radial profile of the on-axis PSF is $\sim10^{\text -4}$ where the calibration speckles land (\autoref{fig:onsky_psf_profiles}). If the speckle contrast is $<10^{\text{-}2}$, then the astrogrid will not affect contrast outside of the speckle PSF cores. A relative brightness of $10^{\text{-}2}$ corresponds to an amplitude of \SI{24}{\nano\meter} for the 10$\lambda/D$ pattern, \SI{28}{\nano\meter} for the 15.5$\lambda/D$ pattern, and \SI{103}{\nano\meter} for the 31$\lambda/D$ pattern.

There are practical considerations for using astrogrid simultaneously with other SCExAO modules. For example, when using SCExAO/CHARIS for PDI, the field stop requires using the 10.3$\lambda/D$ separation grid so that all the speckles land within the CHARIS FOV. The speckle brightness in the NIR is fainter, which means the optimal brightness for CHARIS data might produce brighter than optimal speckles in VAMPIRES. Lastly, for observations in the Open filter, the radial smearing of the satellite spots reduces the astrometric and photometric precision.

\begin{deluxetable}{lccc}
\tabletypesize{\small}
\tablehead{
    \multirow{2}{*}{Pattern} & \colhead{Separation} & \multirow{2}{*}{$c_1$} & \multirow{2}{*}{$c_0$} \\
                             & \colhead{($\lambda/D$)} & & 
}
\tablecaption{Astrogrid relative photometry flux scaling.\label{tbl:astrogrid}}
\startdata
XYgrid & 10.3 & \num{14.217+-1.489} & \num{-0.111+-0.093} \\
XYgrid & 15.5 & \num{7.634+-1.321} & \num{-0.043+-0.082} \\
XYgrid & 31.0 & \num{0.561+-0.256} & \num{-0.002+-0.023} \\
\enddata
\tablecomments{Photometry measured using elliptical apertures with local background subtraction. Astrogrid applied using \SI{1}{\kilo\hertz} modulation speed.}
\end{deluxetable}

\subsection{On-Sky Contrast Curves}

Contrast curves were measured for the CLC-3 and CLC-5 coronagraphs during commissioning. The CLC-3 contrast curve is discussed in \autoref{sec:hd1160}. The CLC-5 contrast was measured from a \SI{60}{\minute} sequence of HD 102438 ($m_{\rm R}=$6.0, \citealp{zacharias_ucac5_2017}) with $\sim$\ang{10} of field rotation. The conditions were good (\ang{;;0.45}$\pm$\ang{;;0.1} seeing), but LWE caused frequent PSF splitting.

\begin{figure}
    \centering
    \script{contrast_curves_HD102438.py}
    \includegraphics{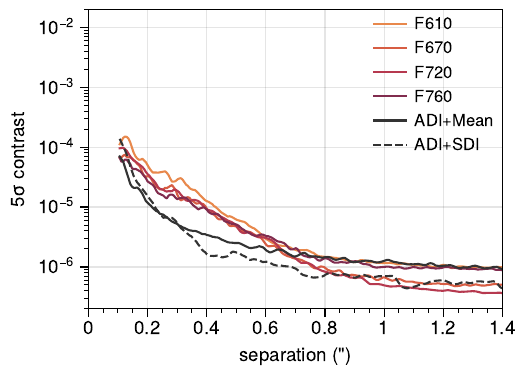}
    \caption{5$\sigma$ throughput-corrected Student-t contrast curves from multiband imaging of HD 102438 with the CLC-5 coronagraph. This was \SI{60}{\minute} of data with \ang{10} of field rotation. The contrast from the PCA (20 components) PSF subtraction for each MBI filter is shown with red curves. The solid black curve is the mean-combined residual contrast from each wavelength, and the dashed line is the median SDI-subtracted contrast curve. \label{fig:contrast_curves}}
\end{figure}

The data was post-processed with \texttt{ADI.jl} \citep{lucas_adijl_2020} using principal component analysis (PCA, or KLIP; \citealt{soummer_detection_2012}) with 20 principal components for each multiband field. These results produce residual frames, which are then derotated and collapsed with a weighted mean \citep{bottom_noise-weighted_2017}. The contrast was measured by dividing the noise in concentric annuli by the stellar flux, correcting for small sample statistics \citep{mawet_fundamental_2014}. Algorithmic throughput was measured by injecting off-axis sources and measuring the recovered flux after PSF subtraction. The contrast curves for each filter are shown in \autoref{fig:contrast_curves}.

The best contrast is achieved at the longest wavelengths, following the inverse wavelength scaling of optical path differences. There is a slight deviation from this behavior in the F760 filter at far separations where the curve becomes background-limited, which can be attributed to the lower average S/N in the truncated bandpass of the F760 filter (\autoref{tbl:filters}). Roughly $10^{\text{-}4}$ contrast was reached at the IWA (\ang{;;0.1}), $10^{\text{-}5}$ at \ang{;;0.4}, and $10^{\text{-}6}$ beyond \ang{;;0.6}.

\subsection{Spectral Differential Imaging}\label{sec:sdi}

The wavelength diversity of multiband imaging data was leveraged by performing spectral differential imaging (SDI) in two ways. The first was by taking each collapsed ADI residual frame and doing a mean combination of the residuals, attenuating uncorrelated noise (e.g., read noise). The second technique uses another round of PSF subtraction after radially scaling each frame by wavelength and multiplying by a flux factor (``ADI + SDI''). The median PSF from the scaled images was subtracted from the data before rescaling back to sky coordinates (``classical'' SDI).

Fake companions were injected with the same spectrum as the star at equal contrasts for each wavelength for algorithmic throughput analysis. 5$\sigma$ contrasts of $10^{\text{-}4}$ at were reached at the IWA (\ang{;;0.1}), $10^{\text{-}5}$ contrast at \ang{;;0.2}, and $10^{\text{-}6}$ beyond \ang{;;0.5}. In general, the SDI reduction achieves better contrast than individual ADI reductions alone, especially from \ang{;;0.1} to \ang{;;0.7}.

\subsection{Residual Atmospheric Dispersion}

\begin{figure}
    \centering
    \script{resid_adc.py}
    \includegraphics{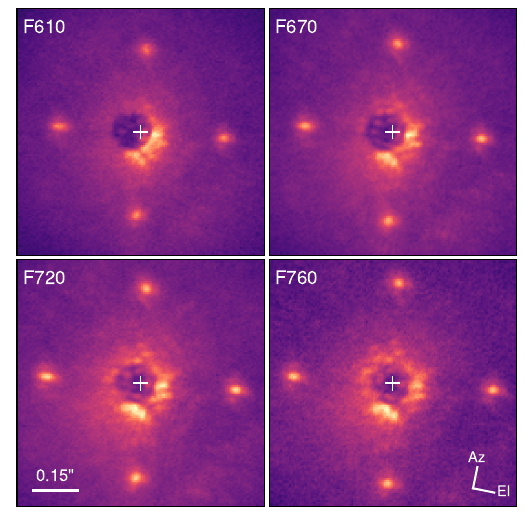}
    \caption{Coronagraphic multiband imaging with the CLC-3 coronagraph at 2.25 airmass (\ang{26.3} elevation). The filter is labeled at the top left, the apparent location of the star is marked with a cross, and the telescope elevation and azimuth axes are displayed with a compass at the bottom left. The image is shown with a square-root stretch with separate limits for each filter. The residual atmospheric dispersion differentially shifts the star in the focal plane, reducing the effectiveness of the coronagraphic diffraction control.\label{fig:resid_adc}}
\end{figure}

There is only one broadband atmospheric dispersion corrector (ADC) in the common path of SCExAO \citep{egner_atmospheric_2010}, resulting in residual atmospheric dispersion at low elevations. This dispersion causes a differential shift of the PSF at each wavelength, which can be problematic for coronagraphy with multiband imaging because the PSF can leak outside the focal plane mask (\autoref{fig:resid_adc}). With the CLC-3 mask, this effect is noticeable above an airmass of $\sim$\num{2.2} ($<$\ang{27} elevation). The larger masks (CLC-5 and CLC-7) are less affected by this leakage.

\section{Imaging Polarimetry}\label{sec:polarimetry}

Polarimetric differential imaging (PDI; \citealp{kuhn_imaging_2001}) is a powerful technique for high-contrast observations. By measuring orthogonal polarization states simultaneously through nearly the same optical path, the unpolarized signal in the stellar PSF can be canceled to a high degree (less than $\Delta10^{-4}$, \citealp{schmid_spherezimpol_2018}). This differential approach significantly suppresses the stellar PSF, allowing the detection of faint polarized signals from circumstellar disks or exoplanets. Additionally, through polarimetric modulation, it is possible to remove or greatly reduce the effects of instrumental polarization and fast atmospheric seeing.
 
\subsection{Polarimetric Differential Imaging}

VAMPIRES is designed to measure linear polarization with high precision and accuracy. A polarizing beamsplitter cube splits horizontally polarized light ($I_{0^\circ}$) to VCAM1 and vertically polarized light ($I_{90^\circ}$) to VCAM2. This intensity mixes the inherently polarized signal and the inherently unpolarized signal ($I_\star$)
\begin{equation}
    I_{C1} = I_{0^\circ} + 0.5\cdot I_\star,
\end{equation}
\begin{equation}
    I_{C2} = I_{90^\circ} + 0.5\cdot I_\star.
\end{equation}

Subtracting data from the two cameras yields the Stokes $Q$ parameter, by definition--
\begin{equation}
    I_{C1} - I_{C2} = I_{0^\circ} - I_{90^\circ} + 0.5\cdot \left( I_\star - I_\star\right) = Q.
\end{equation}
Conveniently, the unpolarized stellar signal is completely removed by this subtraction.

To measure Stokes $U$, which is the difference of light at \ang{45} and \ang{135}--
\begin{equation}
    U = I_{45^\circ} - I_{135^\circ}.
\end{equation}
The input signal is modulated with a HWP so that $I_{45^\circ}$ and $I_{135^\circ}$ are retarded into horizontally and vertically polarized light, which then can be measured by the cameras.

In theory, the subtraction in the definition of the Stokes observables removes the unpolarized signal, but, in reality, the imperfections of the polarimeter leave instrumental polarization and non-common path errors \citep{kuhn_imaging_2001,tinbergen_astronomical_2005}. Every inclined reflective surface will partially polarize otherwise unpolarized light, and these reflections are difficult to avoid in a complicated instrument like SCExAO mounted on a Nasymth platform \citep{tinbergen_accurate_2007}. The instrumental polarization is a nuisance term in the analysis, but a majority of it is canceled by the HWP modulation-- rotating the HWP \ang{45} switches the sign of the observable (Stokes $Q$ becomes $-Q$). The difference ($[Q - (-Q)]/2$) removes the instrumental polarization downstream of the HWP.

\subsection{High-Speed Polarimetric Modulation}

Switching orthogonal polarization states at the same rate or faster than the atmospheric speckle timescale ($\sim$\si{\milli\second}) differentially removes these fast speckles with high precision during the polarimetric data reduction \citep{kemp_piezo-optical_1969}. This fast modulation technique has been successfully implemented on several aperture polarimeters \citep{rodenhuis_extreme_2012,harrington_innopol_2014,bailey_high-sensitivity_2015,bailey_high-precision_2017,bailey_hippi-2_2020}, seeing-limited polarimeters \citep{safonov_speckle_2017,bailey_picsarr_2023}, and high-contrast polarimeters \citep{norris_vampires_2015,schmid_spherezimpol_2018}. 

VAMPIRES uses a ferroelectric liquid crystal (FLC) for fast polarimetric modulation. The FLC behaves like a HWP whose fast-axis rotates \ang{45} when a voltage is applied. VAMPIRES was upgraded with an achromatic FLC rotator\footnote{Meadowlark FPA-200-700}, which has better retardance across the bandpass than the previous one. The initial characterization of the FLC shows a $\sim$20\% improvement in the efficiency of measuring the linear Stokes parameters.

The FLC switches in less than \SI{100}{\micro\second}, which is synchronized with every detector exposure (\SI{1}{\hertz} to $>$\SI{1}{\kilo\hertz}). The new FLC is mounted in a temperature-controlled optical tube at \SI{45}{\celsius} for consistent, efficient modulation and to avoid any temperature-dependent birefringence of the liquid crystal. The FLC photometric throughput is 96.8\%. The FLC has a maximum excitation time of \SI{1}{\second}, which means it can only be used with short exposures. The FLC mount was motorized to allow removing it for ``slow'' polarimetry, which only uses the HWP for modulation.

\subsection{Instrumental Modeling and Correction}

Polarimetric measurements are limited by instrumental polarization due to the many inclined surfaces and reflections in the common path of VAMPIRES, in particular from M3 \citep{tinbergen_accurate_2007}. Mueller calculus is a well-established technique to model and remove the instrumental effects \citep{perrin_polarimetry_2015,holstein_polarimetric_2020,joost_t_hart_full_2021}. The Stokes parameters measured through difference imaging are a measurement of the input Stokes values modified by the telescope and instrument Mueller matrix--
\begin{equation}
    \mathbf{S}(x, y) = \mathbf{M}\cdot\mathbf{S}_{in}(x, y),
    \label{eqn:mm}
\end{equation}
where $\mathbf{S}$ represents Stokes vector at each point in the field. With a calibrated model of $\mathbf{M}$, the instrument Mueller-matrix, \autoref{eqn:mm} can be solved using a least-squares fit to estimate the input Stokes vector \citep{perrin_polarimetry_2015}.

For VAMPIRES the Mueller matrix, $\mathbf{M}$, is estimated with a forward model. This model includes offset angles, diattenuations, and retardances for the polarizing components in each filter \citep{zhang_characterizing_2023}. The model is fit to calibration data with a polarized flat source-- these calibrations are automated for consistent monitoring of the instrument's polarimetric performance. 

The telescope tertiary mirror is a significant source of instrument polarization due to the small focal ratio (F/13.9) and \ang{45} inclination \citep{schmid_spherezimpol_2018,van_holstein_polarization-dependent_2023}. Because the internal polarized source is after M3 (\autoref{fig:schematic}), these birefringent effects cannot be easily calibrated without on-sky observations. A full instrument characterization and measurements of unpolarized ($p<$0.01\%) and polarized ($p>$1\%) standard stars to calibrate the effects of M3 is future work \citep{zhang_characterizing_2023}. Observers needing high-precision polarimetry should observe an unpolarized and a polarized standard star: at least one HWP cycle at the beginning and end of the science sequence.

\section{Polarimetric Differential Interferometry}\label{sec:interferometry}

\begin{figure*}[t]
    \centering
    \script{nrm_masks.py}
    \includegraphics[width=\textwidth]{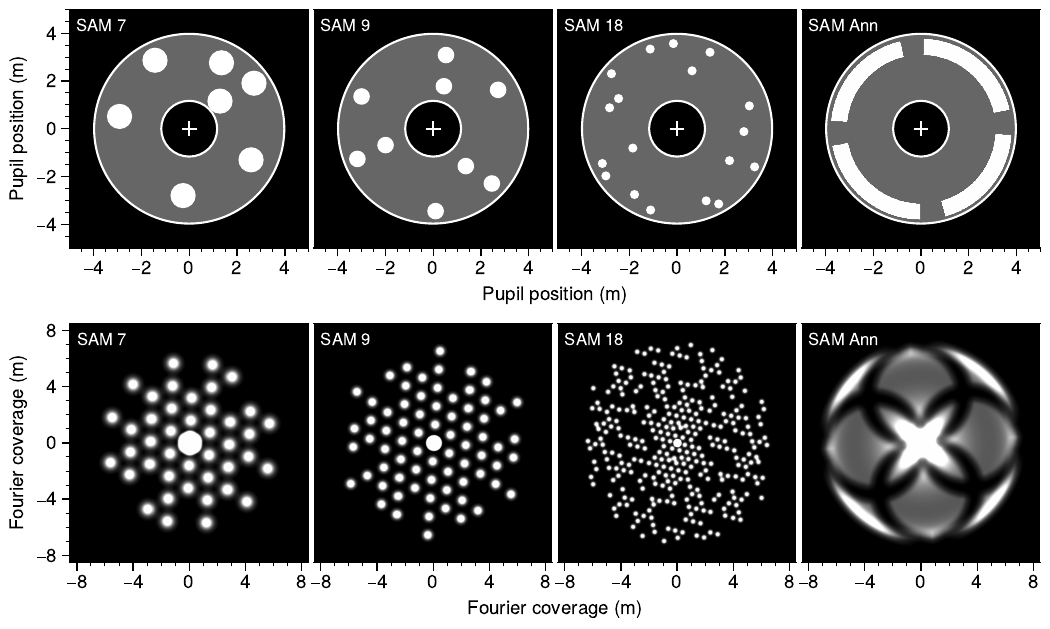}
    \caption{The four aperture masks available for use with the VAMPIRES instrument. (Top) design of masks in the pupil plane. The masks are named according to the number of sub-apertures. (Bottom) u,v-plane Fourier coverage. The choice of the mask is a function of the required Fourier coverage and the required throughput for a high signal-to-noise measurement of a target. \label{fig:nrm_masks}}
\end{figure*}

VAMPIRES features a single telescope interferometric mode through non-redundant aperture masking (NRM). NRM is implemented by placing an opaque mask with a sparse array of sub-apertures in the pupil plane of the instrument. Four masks are available for use with VAMPIRES (\autoref{fig:nrm_masks}).

The non-redundant spacing of these sub-apertures, wherein each has a unique vector separation, ensures unique measurements of each baseline length (Fourier component) of the astrophysical scene \citep{tuthill_aperture_2000}. Unambiguous measurements of each Fourier component yield observables that are robust to atmospheric seeing and, as such, do not contain ``redundancy'' noise that grows with the square root of the number of measurements. As a result, NRM allows for the reconstruction of information at angular resolutions at and beyond the diffraction limit \citep{labeyrie_introduction_2014}. Combining the advantages of NRM with those of polarimetry (\autoref{sec:design}, \autoref{sec:polarimetry}) enables sub-diffraction limited measurement of polarized signal within the previously inaccessible regions of the image. In particular, the NRM mode unlocks regions obscured within the coronagraphic masks' inner-working angle (\autoref{sec:coronagraphy}). In this way, the NRM and coronagraphic modes operate in a complementary fashion-- the outer working angle of the NRM mode is approximately the inner working angle of the coronagraph \citep{norris_vampires_2015}, and together, these modes provide near complete coverage of the circumstellar environment, right down to the stellar surface. For an appropriate choice of target (notably, bright enough to counteract the throughput loss of the mask), the NRM mode of VAMPIRES provides a powerful characterization of polarized circumstellar environments, limited only by contrast.

\begin{figure}[h]
\centering
    \script{nrm_visibilities.py}
    \includegraphics[width=\columnwidth]{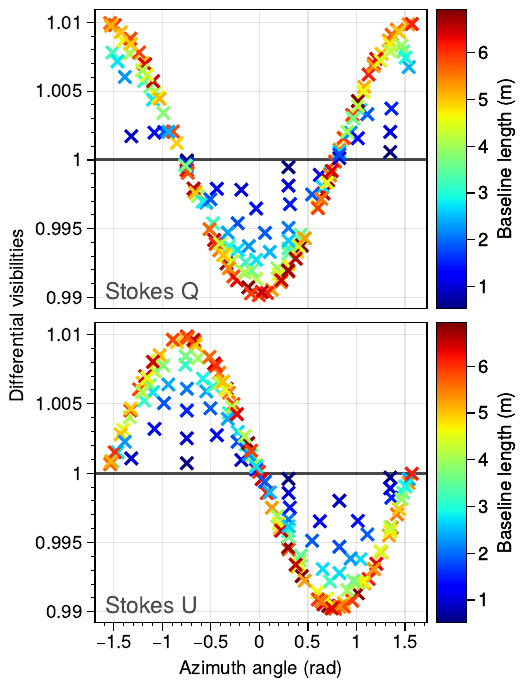}
    \caption{Polarimetric, differential visibilities for a model of a spherically symmetric circumstellar shell. Visibilities for a symmetric source have a distinctive sinusoidal appearance. Differential visibilities are produced for Stokes Q (top) and Stokes U (bottom), which describe the complete characterization of the linearly polarised circumstellar environment. \label{fig:diff_vis_all}}
\end{figure}

The interferometric observables are the differential, polarimetric analogs of standard visibilities and closure phases. They are self-calibrated (no calibrator star is required, and errors induced by AO are removed) and relatively immune to non-common path instrumental noise \citep{norris_vampires_2015}. Raw data is reduced to visibilities and closure phases using the VAMPIRES branch of the AMICAL package \citep{soulain_james_2020} and further reduced to differential visibilities and closure phases using our pipeline (Lilley et al., 2024). Models and images of the astrophysical scene can then be fit to these resulting observables. Techniques for model fitting and novel machine learning-based image reconstruction techniques are currently in development. They will be released soon, along with early science results obtained using the NRM mode (Lilley et al., 2024). 
 
The VAMPIRES upgrades provide several key improvements for NRM. First, the MBI (\autoref{sec:mbi}) mode enables simultaneous measurement of differential observables (e.g., spectral differential phase), allowing for the high-precision characterization of circumstellar dust grain size and species. This improves observing efficiency by a factor of four, compared to repeating observations for different filters. Since the time-varying wavefront error is simultaneously sampled across wavelengths, direct interferometric phase measurements between spectral components can be produced. The increased observing efficiency enables the observer to accumulate more parallactic angle coverage, which is critical for characterizing non-azimuthal scattering symmetries. The new qCMOS detectors have a higher dynamic range and faster framerate than the previous EMCCDs \autoref{sec:detectors}, which creates sharper fringes without saturation for sufficiently bright targets. The qCMOS detectors also have better sensitivity in the photon-starved regime, as they do not suffer from the excess noise of electron multiplication from which EMCCDs suffer (\autoref{fig:detector_snr_relative}). 
The development and characterization of the NRM mode are ongoing, and a comprehensive analysis will be published in the near future (Lilley et al., 2024).

\section{Data processing}\label{sec:processing}

A purpose-built open-source data processing pipeline for VAMPIRES is under development.\footnote{\url{https://github.com/scexao-org/vampires_dpp}} This pipeline is designed to reduce data for both versions of the instrument and focuses on frame calibration, alignment, and polarimetry. The pipeline works for all observing modes, except non-redundant masking, and produces calibrated and coadded data ready for post-processing with other tools.

\subsection{Pre-processing}
For pre-processing the data is organized by type and object. Master calibration files are prepared from dark frames, sky frames, and flat frames. Multiband flats require an additional step to fit and normalize each multiband field independently. The final pre-processing step prompts the user to click on the approximate center of the field of view to aid in registration.

\subsection{Frame Calibration and Collapsing}

Raw data is calibrated with dark subtraction, flat-field normalization, and bad-pixel correction. Data is converted into units of \si{\electron/\second} and the FITS headers are normalized across the different instrument versions. Precise proper motion-corrected coordinates and parallactic angles are calculated from GAIA astrometry, when available \citep{gaia_collaboration_gaia_2016,gaia_collaboration_gaia_2018,gaia_collaboration_gaia_2021}.

After calibration, the frame centroid is measured using cross-correlation with a PSF model-- for coronagraphic data, the four satellite spots (\autoref{sec:astrogrid}) are measured independently.  The data is registered with the centroids and PSF statistics like the Strehl ratio, FWHM, and the photometric flux are measured. The individual FOVs in multiband data are analyzed separately and then cut out and stacked into a spectral cube. Absolute flux calibration is supported by providing a stellar spectrum, either a model or a calibrated spectrum, which is used for synthetic photometry to derive the flux conversion factor.

The data is collapsed with optional frame selection for lucky imaging. The registered and collapsed data frames are saved to disk for post-processing with other tools, such as \texttt{ADI.jl} \citep{lucas_adijl_2020}, \texttt{VIP} \citep{gomez_gonzalez_vip_2017,christiaens_vip_2023}, or \texttt{PyKLIP} \citep{wang_pyklip_2015}.

\subsection{Polarimetry}

Polarimetric data is collated into sets for each HWP cycle and Stokes cubes are produced from the double- or triple-difference procedure. Before differencing, data are corrected for small differences in plate scale and angle offset between the cameras. The Stokes cube is optionally corrected with a Mueller-matrix, following \citet{holstein_polarimetric_2020,zhang_characterizing_2023}. Finally, all the individual Stokes cubes are derotated and median-combined. The Stokes $I$, $Q$, and $U$ frames are saved alongside the azimuthal Stokes parameters $Q_\phi$ and $U_\phi$ \citep{monnier_multiple_2019,boer_polarimetric_2020}--
\begin{align}
\begin{split}
    \label{eqn:az_stokes}
    Q_\phi &= -Q\cos{\left(2\phi\right)} - U\sin{\left(2\phi\right)} \\
    U_\phi &= Q\sin{\left(2\phi\right)} - U\cos{\left(2\phi\right)},
\end{split}
\end{align}
where
\begin{equation}
    \phi = \arctan{\left( \frac{x_0 - x}{y - y_0} \right)} + \phi_0
\end{equation}
in image coordinates. Note, this is equivalent to the angle east of north when data has been derotated. The total linear polarization,
\begin{equation}
    P = \sqrt{Q^2 + U^2},
\end{equation}
and the angle of linear polarization,
\begin{equation}
    \chi = \frac12\arctan{\frac{U}{Q}},
\end{equation}
are also calculated and saved in the output cube.
\section{Commissioning Results}\label{sec:firstlight}

The VAMPIRES upgrades were tested on-sky from June to August 2023. Some results from the commissioning observations are presented below. During testing, many of the instrument configurations (polarimetry, coronagraphy, multiband imaging, narrowband imaging) were explored to verify that VAMPIRES is ready for observers. All data were obtained using the 188-mode correction from AO188 and $\sim$1,100 corrected modes, on average, with SCExAO.

\subsection{Spectro-Polarimetric Imaging of HD 169142\label{sec:hd169142}}

VAMPIRES spectro-polarimetric capabilities are highlighted with multiband imaging of the young star \object{HD 169142}. This system hosts a protoplanetary disk \citep{quanz_gaps_2013} and emission signatures from potential protoplanets \citep{hammond_confirmation_2023}. The disk itself is nearly face-on ($i{\sim}$\ang{13}), and includes two rings ($R_{\rm in}{\sim}$\SI{170}{\mas}, $R_{\rm out}{\sim}$\SI{720}{\mas}), with a gap between them. A cavity is close to the star within the inner ring \citep{fedele_alma_2017}.

This system has been imaged in polarized light by \citet{kuhn_imaging_2001,quanz_gaps_2013,monnier_polarized_2017,pohl_circumstellar_2017}, including in the visible with ZIMPOL by \citet{bertrang_hd_2018}. The polarized images match the radio continuum disk morphology, comprising a bright inner ring with a fainter outer ring. The disk is nearly face-on, with an estimated inclination close to \ang{13} \citep{fedele_alma_2017}.

HD 169142 was observed with VAMPIRES in the ``slow'' polarimetry mode (no FLC) with the \SI{105}{\mas} Lyot coronagraph for \SI{55}{minutes}, resulting in \SI{47}{minutes} of data-- roughly an 85\% observing efficiency, including all overheads from external triggering and HWP rotation. The average seeing for the night from the Maunakea weather center (MKWC) archive\footnote{\url{http://mkwc.ifa.hawaii.edu/current/seeing/index.cgi}} was \SI{0.8\pm0.2}{''}.

The data was reduced as described in \autoref{sec:processing} with background subtraction and double-difference polarimetric reduction. Mueller matrices were adapted from \citet{zhang_characterizing_2023} for the common-path HWP, image rotator, and generic optics term (i.e., excluding the detector intensity ratio and the FLC, since these components were modified in the upgrade). Residual instrumental polarization was corrected using the photometric sums in an annulus from \ang{;;0.24} to \ang{;;0.33}, which matches the cavity of the disk \citep{bertrang_hd_2018}.

\begin{figure*}[t]
    \centering
    \script{HD169142.py}
    \includegraphics[width=0.9\textwidth]{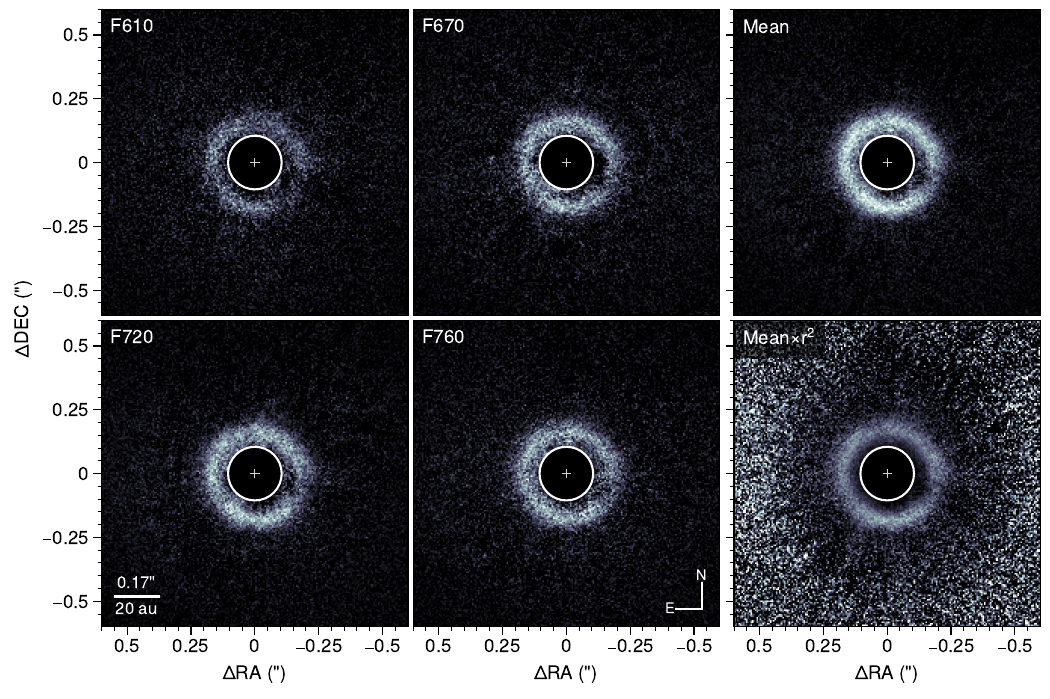}
    \caption{VAMPIRES polarimetric observations of HD 169142 in multiband imaging mode. The Stokes $Q_\phi$ frames are cropped to the inner \ang{;;1.2} FOV with the coronagraph mask marked with a black circle. The data from each multiband filter is shown in the left four quadrants, and the wavelength-collapsed data is shown in the right column. The bottom-right image is multiplied by the squared stellocentric distance ($r^2$), which better shows the edge of the ring at $\sim$\ang{;;0.4}. All data are rotated north up and east left and shown in a linear scale with separate limits for all images.\label{fig:hd169142_mosaic}}
\end{figure*}

The Stokes $Q_\phi$ frames, which are approximately equal to the polarized intensity, are shown in \autoref{fig:hd169142_mosaic}. The inner disk is evident, with some diminished intensity to the southwest. A version scaled by the squared stellocentric distance to normalize the stellar irradiation is also shown, highlighting the disk gap and the inner rim of the outer disk.

\begin{figure}[h]
    \centering
    \script{HD169142.py}
    \includegraphics[width=\columnwidth]{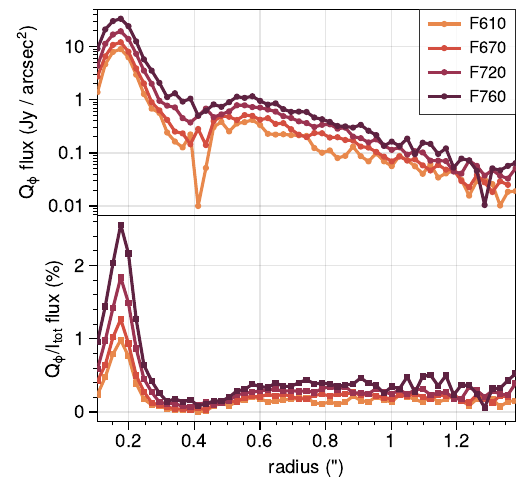}
    \caption{Radial profiles from multiband polarimetric images of HD 169142 for the polarized intensity (top; taken from $Q_\phi$ frame) and fractional polarized flux (bottom; $Q_\phi/I$). The profiles are taken with annuli of a width of 4 pixels starting from the coronagraph IWA (\si{105}{mas}). The region of the inner disk shows clearly that there is more dust scattering at redder wavelengths.\label{fig:hd169142_flux}}
\end{figure}

Radial profiles with bin sizes of 4 pixels from each $Q_\phi$ frame and each Stokes $I$ frame were measured from the coronagraph IWA (\ang{;;0.1} out to \ang{;;1.4}). The ratio of these profiles is shown alongside the $Q_\phi$ profiles in \autoref{fig:hd169142_flux}. The mean integrated partial-polarization of the disk is \SI{0.6}{\%} across all wavelengths and the scattering intensity is higher at redder wavelengths. 

Future work includes modeling geometric disk profiles to analyze the scattering geometry of this disk, allowing a measurement of the scattering phase function of the dust at multiple wavelengths. These measurements give color information about the scattering properties of the dust, which generally requires multiple observations for each filter or integral-field spectroscopy, but with VAMPIRES, they are obtained in one observing sequence using multiband imaging.

\subsection{Imaging of HD 1160B\label{sec:hd1160}}

\object{HD 1160} is an A0V star with two substellar companions \citep{nielsen_gemini_2012,maire_first_2016,garcia_scexao_2017,mesa_characterizing_2020}. The inner companion, HD 1160B, is an M5-M7 brown dwarf with \ang{;;0.7} separation on a \SIrange{252}{1627}{yr} orbit \citep{blunt_orbits_2017}. HD 1160 was observed on \formatdate{11}{7}{2023} using VAMPIRES multiband imaging with a \SI{59}{\mas} IWA coronagraph for a relatively short sequence (17 minutes, \ang{14} PA rotation). The data was processed with light frame selection (discarded the worst 25\% of frames) and flux calibrated using an A0V stellar model. HD 1160B was not detected in the calibrated frames, alone.

\begin{figure*}[t]
    \centering
    \script{HD1160_mosaic.py}
    \includegraphics[width=0.9\textwidth]{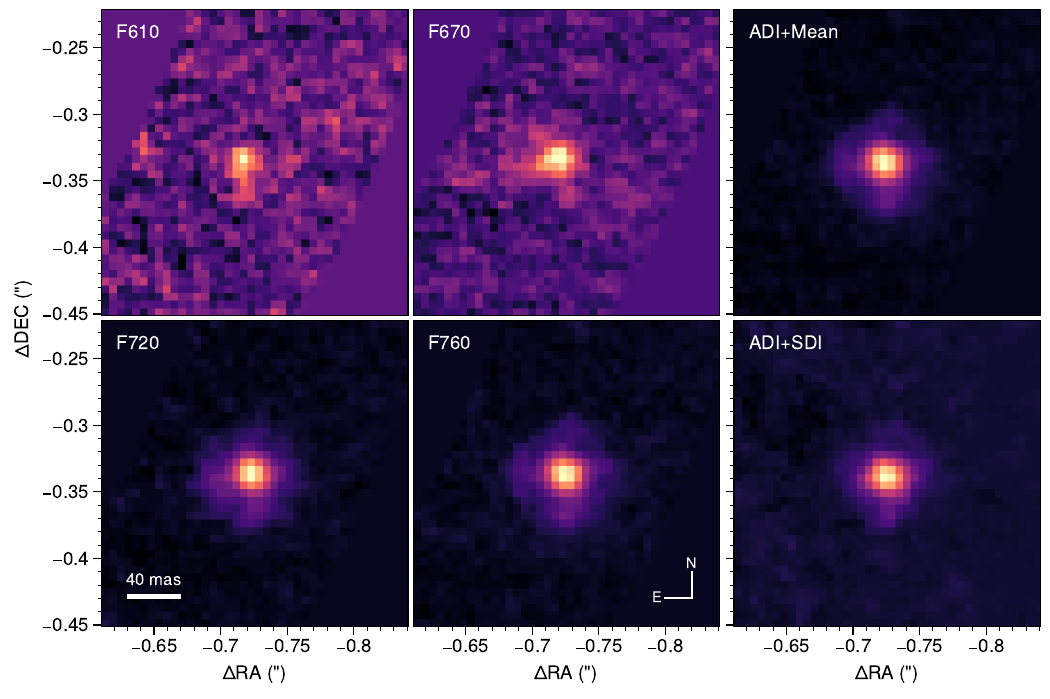}
    \caption{ADI residual frames from VAMPIRES observations of HD 1160 zoomed into a \SI{40}{\pixel}-crop around the companion HD 1160B. Data are shown with a linear scale and different limits for each frame. All frames were processed using the GreeDS algorithm with 20 principal components. The left four frames are residuals from each multiband filter. The top-right frame is the wavelength-collapsed residual, and the bottom-right frame is the ADI+SDI residual which includes a median PSF subtraction in the spectral domain. The ADI+SDI residual has a radial subtraction signature pointing towards the host star due to SDI PSF subtraction.\label{fig:hd1160_mosaic}}
\end{figure*}

\begin{figure}[h]
    \centering
    \script{contrast_curves_HD1160.py}
    \includegraphics[width=\columnwidth]{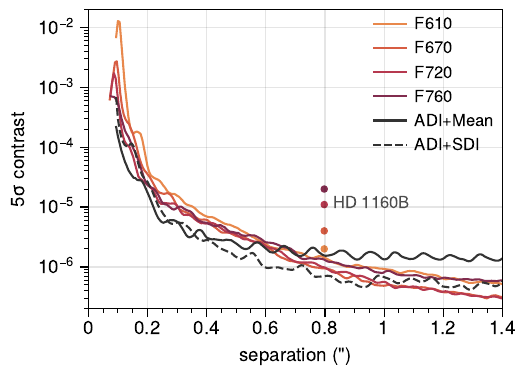}
    \caption{5$\sigma$ throughput-corrected Student-t contrast curves from multiband imaging of HD 1160 with the CLC-3 coronagraph. This was \SI{16}{\minute} of data with \ang{14.4} of field rotation. The contrast from the GreeDS (20 components) PSF subtraction for each MBI filter is shown with red curves. The solid black curve is the mean-combined residual contrast from each wavelength, and the dashed line is the median SDI-subtracted contrast curve. The contrast of HD 1160B at each wavelength is shown with points.\label{fig:hd1160_contrast}}
\end{figure}

PSF subtraction was performed at each wavelength using an iterative ADI algorithm (GreeDS; \citealt{pairet_reference-less_2019,pairet_mayonnaise_2020,stapper_iterative_2022}). The companion was easily detected using 20 principal components. The ADI residuals were combined with and without SDI (\autoref{sec:sdi}). The residual frames are shown in \autoref{fig:hd1160_mosaic}. The throughput-corrected contrast curves (\autoref{fig:hd1160_contrast}) were measured following the same procedure as \autoref{sec:coronagraphy}, except with a mask for the companion to avoid bias.

To determine the precise astrometry of the companion, \texttt{KLIP-FM} \citep{wang_pyklip_2015} was used to forward model a synthetic PSF processed by the KLIP algorithm \citep{pueyo_detection_2016}. For this modeling, 50 principal components were used for each wavelength before mean-collapsing along the spectral axis. The forward-modeled PSF was used to perform an MCMC fit for the position of the companion in the residual frame. The 95\% highest posterior density for the separation was \SI{798.6\pm 1.9}{\mas} at a position angle of \ang{247.07\pm 0.15}. This measurement was used for the astrometric calibration in \autoref{sec:detectors}.

\subsection{Narrowband Imaging of the R Aqr Nebula\label{sec:raqr}}

VAMPIRES high angular resolution, dynamic range, and sensitivity are highlighted by emission line imaging of the innermost regions of the jets and binary of \object{R Aqr}. R Aqr is a symbiotic binary-- the primary component is an AGB star (a Mira variable) and the secondary is a white dwarf (WD) \citep{merrill_partial_1935,merrill_spectra_1940}. Due to the proximity of the WD (\SIrange{10}{50}{\mas} over the orbit), matter is transferred directly from the giant star onto a compact accretion disk around the WD, creating a bipolar jet that extends $\sim$\ang{;;30} to the NE and SW \citep[and references therein]{schmid_spherezimpol_2017}. The \SI{42}{yr} orbit of the binary has been studied in detail using radial velocity measurements, and recently with visual ephemerides from visible and radio imaging \citep{gromadzki_spectroscopic_2009,bujarrabal_high-resolution_2018,alcolea_determining_2023}. 

The WD companion has not been detected directly in the visible/NIR; at radio wavelengths, the CO continuum can be mapped with sufficient angular resolution to resolve the WD \citep{bujarrabal_high-resolution_2018,alcolea_determining_2023}. \citet{schmid_spherezimpol_2017} showed using ZIMPOL that there is a bright H$\alpha$ emission feature coincident with the WD, thought to be a compact accretion disk and the origin of the bipolar jet. In 2020, a periastron passage occurred, putting the WD only a few \si{au} from the AGB star. This periastron passage was studied at many wavelengths \citep{hinkle_2020_2022,sacchi_front-row_2024}.

\begin{figure}[h]
    \centering
    \script{RAqr.py}
    \includegraphics[width=\columnwidth]{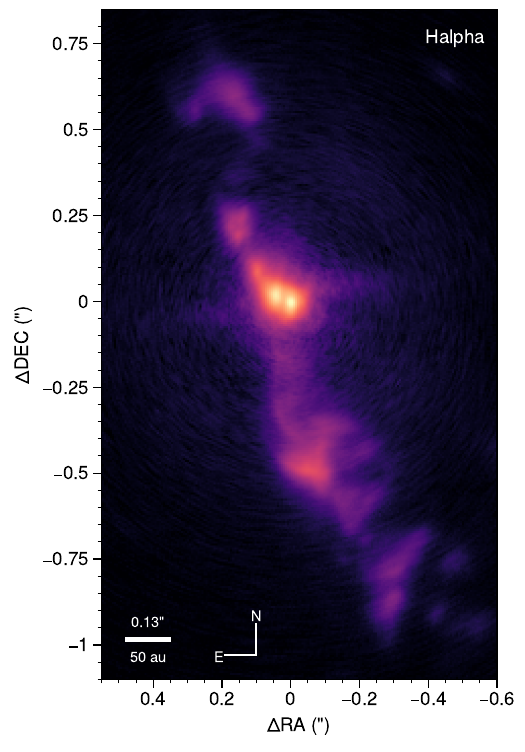}
    \caption{VAMPIRES H$\alpha$ narrowband image of R Aqr with logarithmic stretch showing the emission nebula. A faint diffraction spike, which is not astrophysical, can be seen roughly aligned with the RA axis around the central source.\label{fig:raqr}}
\end{figure}

\begin{figure}[h]
    \centering
    \script{RAqr.py}
    \includegraphics[width=\columnwidth]{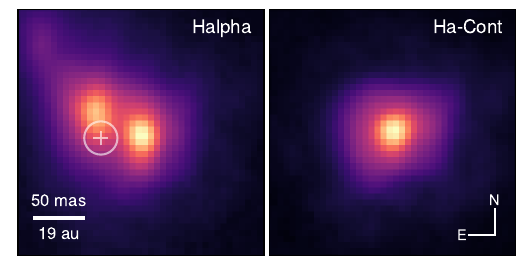}
    \caption{A zoom-in of the inner \ang{;;0.24} FOV of R Aqr. (left) H$\alpha$ narrowband image shows the AGB star and a bright jet emanating from the WD. The expected location of the WD from the orbit of \citet{alcolea_determining_2023} is marked with a white circle one FWHM in width. (right) H$\alpha$-continuum image, which only contains emission from the AGB star.\label{fig:raqr_mosaic}}
\end{figure}

R Aqr was observed with VAMPIRES on \formatdate{7}{7}{2023} using the narrowband H$\alpha$/H$\alpha$-continuum filter pair. Relatively long exposures were used (\SI{1.2}{s}), which blurred the PSF and reduced the efficacy of lucky imaging (\autoref{fig:lucky_imaging}). The seeing during the observation was \ang{;;0.8}$\pm$\ang{;;0.2} from the MKWC DIMM data, and the FWHM in the continuum image was \SI{32}{\mas} with a Strehl ratio of 15\%. The collapsed H$\alpha$ image is shown in \autoref{fig:raqr}, and a side-by-side of the inner region with the continuum is shown in \autoref{fig:raqr_mosaic}.

The data shows a resolved PSF-like clump to the northeast and diffuse emission extending from NE to SW. A separation of \SI{50.0}{\mas} at a position angle of \ang{65.0} east of north was measured using PSF models. 

Interestingly, this does not match the observations of \citet{bujarrabal_high-resolution_2018,alcolea_determining_2023} (\autoref{fig:raqr_mosaic}), the H$\alpha$ emission clump is no longer co-located with the WD companion. Unpublished high-resolution ALMA CO observations from a month after these observations show that the WD now has an asymmetric jet following its periastron passage (Alcolea,~J., priv. communication). The location of the asymmetric jet is consistent with the observed H$\alpha$ emission north of the actual position of the WD. Future H$\alpha$ observations tracking the evolution of the jet emission are planned and will provide insight into the dynamical evolution of the R Aqr system.

\subsection{Spectro-Polarimetric Imaging of Neptune\label{sec:neptune}}

\begin{figure*}[t]
    \centering
    \script{neptune.py}
    \includegraphics[width=\textwidth]{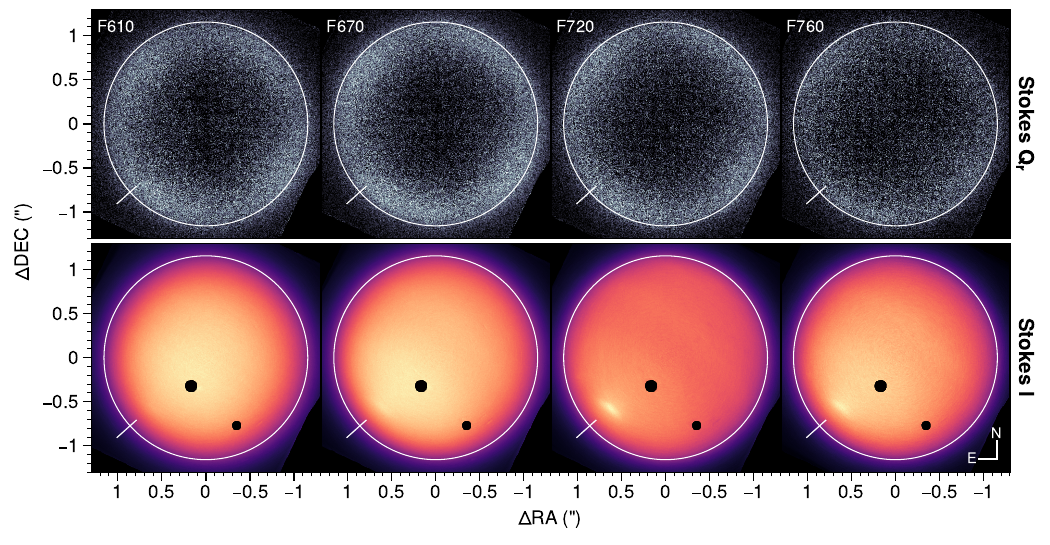}
    \caption{Multiband VAMPIRES observations of Neptune. All data are rotated so to north up and east left in linear scale with different limits for each frame. The multiband filters are shown in each column. The top row is the Stokes $Q_r$ image. The bottom row is Stokes $I$ (total intensity). The apparent diameter of the disk is shown with a circle and the southern polar axis is designated with a line. Flat-field errors in the total intensity images are masked out.\label{fig:neptune_mosaic}}
\end{figure*}

\object{Neptune} is a compelling target for polarimetric imaging due to its dynamic gaseous atmosphere, which presents a unique opportunity to study scattering processes distinct from those in circumstellar disks (where single-scattering events dominate). Neptune has previously been imaged with ZIMPOL \citep{schmid_limb_2006} when it was installed on the ESO \SI{3.6}{\meter} telescope. 

In this work, we present high-resolution spectro-polarimetric observations of Neptune with VAMPIRES, taken on \formatdate{11}{7}{2023} in slow polarimetry mode. The angular diameter of Neptune (\ang{;;2.3}) sat comfortably within the VAMPIRES FOV, eliminating the need for dithering. The AO performance was hindered due to the lack of a clear point source to use as a natural guide star. The planetary ephemeris from JPL horizons\footnote{\url{https://ssd.jpl.nasa.gov/horizons/}} is shown in \autoref{tbl:neptune}. 

\begin{deluxetable}{ll}
\tablewidth{\columnwidth}
\tablehead{\colhead{Parameter} & \colhead{Value}}
\tablecaption{Neptune ephemeris for \formatdate{11}{7}{2023} based on JPL horizons data.\label{tbl:neptune}}
\startdata
Apparent diameter & \ang{;;2.313} \\
North pole angle & \ang{318.1} \\
North pole distance & \ang{;;-1.058} \\
Sun-observer phase angle & \ang{1.817}
\enddata
\end{deluxetable}

For processing, each frame was registered using a cross-correlation with the average image. Calibration factors from \autoref{tbl:filters} were used for absolute flux calibration. Similar to \autoref{sec:hd169142}, the polarimetric reduction used double-differencing with an adapted Mueller-matrix correction. The radial Stokes parameters, $Q_r$ and $U_r$ \citep{schmid_limb_2006}, were calculated using
\begin{align}
    \label{eqn:rad_stokes}
\begin{split}
    Q_r &= Q\cos{\left(2\theta\right)} + U\sin{\left(2\theta\right)} \\
    U_r &= -Q\sin{\left(2\theta\right)} + U\cos{\left(2\theta\right)}.
\end{split}
\end{align}
For an optically thick planetary atmosphere like Neptune's, the most probable alignment of the electric field due to multiple-scattering is radially from the center of the planet to the limb; therefore, $Q_r$ approximately equal to the polarized intensity \citep{schmid_limb_2006}. Similar to \autoref{sec:hd169142}, residual polarization errors were corrected with photometry, awaiting the updated instrument polarization calibration. The corrected radial Stokes $Q_r$ frames and total intensity frames are shown in \autoref{fig:neptune_mosaic}.

\begin{figure}
    \centering
    \script{neptune.py}
    \includegraphics[width=\columnwidth]{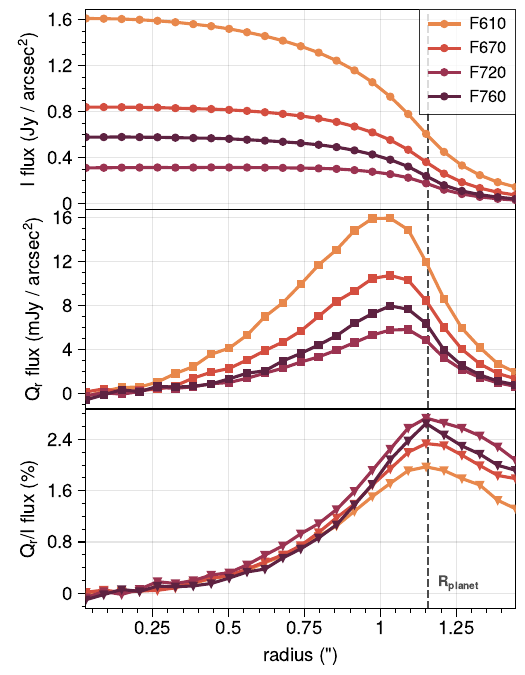}
    \caption{Radial profiles of from spectro-polarimetric observations of Neptune. All profiles use apertures 10 pixels wide. (top) Stokes $I$, total intensity. (middle) radial Stokes $Q_r$. (bottom) the ratio of the $Q_r$ profile to the $I$ profile.\label{fig:neptune_flux}}
\end{figure}

The Stokes $Q_r$ images show clear limb polarization in all filters with little polarization in the center of the image and maximum polarization at the edge. There is azimuthally symmetric polarization along the limb without significant concentrations at the poles or equator. The total intensity frames show signs of a polar cloud and mild banding structures, especially in the F720 frame. Radial profiles of $Q_r$, $I$, and $Q_r/I$ are shown in \autoref{fig:neptune_flux}.

\section{Conclusions}\label{sec:conclusions}

In this paper, we presented upgrades to the visible high-contrast imaging polarimeter VAMPIRES. These upgrades included new detectors, coronagraphs, polarization optics, and a multiband imaging mode. We characterized the instrument and described all current optics and observing modes. A variety of engineering and science validation tests were performed, which confirmed the instrument's readiness.

We highlight key features of VAMPIRES, which emphasize its technological capabilities
\begin{enumerate}
    \item Extreme AO and high-speed lucky imaging makes VAMPIRES capable of visible high-contrast imaging, reaching angular resolutions of $\sim$\SI{20}{\mas} and Strehl ratios up to 60\%.
    \item Two new qCMOS detectors combine high frame rates, low read noise (\SIrange{0.25}{0.4}{\electron} RMS), and large format arrays with subframe readout, no cosmetic defects, and a high dynamic range (\SI{90}{\deci\bel}).
    \item We developed a dichroic-based technique for imaging multiple fields at multiple wavelengths multiplexed across the large detector arrays. This technique is compatible with many instrument modes, effectively increasing the observing efficiency by a factor of four and adding low-resolution spectral information.
    \item We installed a suite of visible coronagraphs and pupil masks, enabling deep observations. Initial tests reached 5$\sigma$ contrast limits of $10^{-4}$ to $10^{-6}$ from \ang{;;0.1} to $>$\ang{;;0.5}.
    \item Advanced polarimetric control allows significant attenuation of unpolarized signal. The new achromatic FLC has improved the polarimetric efficiency by $\sim$20\% and can be removed for long exposures ($>$\SI{1}{\second}). Polarimetry can be incorporated into almost all observational modes of VAMPIRES, including multiband imaging.
\end{enumerate}

With its technological capabilities, VAMPIRES enables the study of many interesting astrophysical phenomena and objects. In \autoref{sec:firstlight} we demonstrated VAMPIRES with 
\begin{enumerate}
\item Spectro-polarimetric imaging of the HD 169142 circumstellar disk, detecting the inner and outer rings and measuring the wavelength dependence of the radial scattering profile
\item High-contrast spectral differential imaging of the sub-stellar companion HD 1160B, measuring its astrometry and spectrum
\item Narrowband H$\alpha$ imaging of the R Aqr emission nebula, resolving the compact accretion source for the jet and discovering a new asymmetry
\item Spectro-polarimetric imaging of gas giant Neptune, confirming the limb polarization of its atmosphere and measuring the wavelength dependence of the radial scattering profile
\end{enumerate}
These results highlight the utility of VAMPIRES as a visible high-contrast instrument.

\subsection{Future Prospects}\label{sec:futureprospects}

The most exciting prospect for VAMPIRES is commissioning the new facility adaptive optics system, AO3k \citep{lozi_ao3000_2022}. This includes upgrading the 188 actuator DM to a 64$\times$64 actuator DM, a NIR pyramid wavefront sensor (NIRWFS), and a visible non-linear curvature wavefront sensor (nlCWFS \citealp{ahn_development_2023}). The AO3k upgrade will enable high-contrast wavefront correction directly from the facility AO,  while the SCExAO PyWFS and DM can act as a second-stage ``touch-up'' loop optimized for small wavefront errors and dark field control.

The NIRWFS is particularly exciting for VAMPIRES since it enables observations of very dusty, red stars. These stars are too faint in the visible for extreme AO but sufficiently bright in the NIR. Circumstellar disk hosts and young stars, in general, are perfect examples of objects optimal for taking advantage of the NIRWFS and VAMPIRES' new long exposure capabilities. High-contrast instruments have poorly studied young and low-mass targets due to their visible faintness, and a new wave of science will be enabled by studying the disks and demographics of these systems.

While \SI{30}{\meter}-class telescopes are still on the horizon, and their HCI instrumentation perhaps even further, instruments like SCExAO and VAMPIRES are one the best places to develop and implement advanced technology, both software and hardware, for the next generation of ground-based high-contrast instruments: ELT/PCS \citep{kasper_pcs_2021}, TMT/PSI \citep{fitzgerald_planetary_2019}, and GMT/GMagAO-X \citep{kautz_gmagao-x_2023}. The multi-wavelength observing techniques and wavefront control methods studied by SCExAO today will directly apply to developing these future instruments, bringing us closer to the goals of direct imaging and characterization of planets most similar to Earth and understanding the formation pathways for planets and solar systems.

\section*{Acknowledgements}
First and foremost, we wish to recognize and acknowledge the significant cultural role and reverence that the summit of Maunakea has always had within the Indigenous Hawaiian community. We are grateful and thank the community for the privilege of building instrumentation on and conducting observations from this sacred Mauna.

We thank Joshua Hacko (NH Micro), Iven Hamilton (UH), Shane Jacobson (UH/Gemini), Michael Lemmin (NAOJ), Paul Sumner (Opto-line), Lucio Ramos (NAOJ), Paul Toyoma (UH), and Matthew Wung (NAOJ) for their time and expertise in the manufacturing, assembly, testing, and integration of the VAMPIRES upgrades.

We thank Javier Alcoloa for the valuable discussions on our R Aqr observations. We thank Timothy Brandt and collaborators for the use of data obtained under Subaru proposal ID o23149. We thank Michael Fitzgerald and collaborators for the use of data obtained under Subaru proposal ID o23150. We thank telescope operators Sean Cunningham, Erin Dailey, Jonathan Laverty, Michael Letawsky, Andrew Neugarten, Larissa Schumacher, and Aidan Walk for their support during the observations used in this work.

This research has made use of the Washington Double Star Catalog maintained at the U.S. Naval Observatory. This research was funded by the Heising-Simons Foundation through grant \#2020-1823. Based on data collected at Subaru Telescope, which is operated by the National Astronomical Observatory of Japan. The development of SCExAO is supported by the Japan Society for the Promotion of Science (Grant-in-Aid for Research \#23340051, \#26220704, \#23103002, \#19H00703, \#19H00695, and \#21H04998), the Subaru Telescope, the National Astronomical Observatory of Japan, the Astrobiology Center of the National Institutes of Natural Sciences, Japan, the Mt Cuba Foundation and the Heising-Simons Foundation.

\facility{Subaru (SCExAO)}
\software{
    \texttt{ADI.jl} \citep{lucas_adijl_2020},
    \texttt{AMICAL} \citep{soulain_james_2020},
    \texttt{astropy} \citep{astropy_collaboration_astropy_2013,astropy_collaboration_astropy_2018},
    \texttt{HCIPy} \citep{por_high_2018},
    Julia \citep{bezanson_julia_2017},
    \texttt{matplotlib} \citep{hunter_matplotlib_2007},
    \texttt{numpy} \citep{harris_array_2020},
    \texttt{pandas} \citep{team_pandas-devpandas_2024},
    \texttt{photutils} \citep{bradley_astropyphotutils_2024},
    \texttt{proplot} \citep{davis_proplot_2021},
    \texttt{pyKLIP} \citep{wang_pyklip_2015},
    Python,
    \texttt{showyourwork} \citep{luger_mapping_2021},
    \texttt{scikit-image} \citep{walt_scikit-image_2014},
    \texttt{synphot} \citep{stsci_development_team_synphot_2018}
}

\bibliography{bib}

\appendix
\section{Observing log\label{sec:log}}

In \autoref{tbl:obslog} we list all observations used for this work.

\begin{deluxetable*}{lllllccccccccc}
\tabletypesize{\footnotesize}
\tablehead{
    \colhead{Date} &
    \multirow{2}{*}{Object} &
    \multirow{2}{*}{Filter(s)} &
    \colhead{Pol.} &
    \multirow{2}{*}{Coron.} &
    \colhead{Pupil} &
    \colhead{DIT} &
    \multirow{2}{*}{$N_\mathrm{coadd}$} &
    \multirow{2}{*}{$N_\mathrm{frame}$} &
    \colhead{$T_\mathrm{exp}$} &
    \colhead{Seeing} &
    \colhead{Prop.} \vspace{-0.75em} \\
    \colhead{(UTC)} & 
    & 
    & 
    \colhead{mode} & 
    &  
    \colhead{mask} & 
    \colhead{(\si{\second})} & 
    & 
    & 
    \colhead{(\si{min})} & 
    \colhead{('')} & 
    \colhead{ID} &
}
\tablecaption{Chronologically-ordered observing log.\label{tbl:obslog}}
\startdata
\formatdate{27}{6}{2023} & Albireo  & 775-50 & - & - & - & \num{5e-4} & 1380 & 2 & $<$1 & - & o23013 \\
\formatdate{27}{6}{2023} & Albireo  & Open & - & - & - & \num{5e-4} & 3450 & 17 & $<$1 & - & o23013 \\
\formatdate{27}{6}{2023} & Albireo  & H$\alpha$/H$\alpha$-Cont & - & - & - & 0.05 & 300 & 12 & 1 & - & o23013 \\
\formatdate{27}{6}{2023} & Albireo & SII/SII-Cont & - & - & - & 0.03 & 313 & 21 & 3 & - & o23013 \\
\formatdate{27}{6}{2023} & Albireo & MBI & - & - & - & \num{2e-3} & 1000 & 20 & 5 & - & o23013 \\
\formatdate{29}{6}{2023} & HD 102438 & MBI & - & CLC-5 & Lyot & 0.01 & 1000 & 383 & 64 & \num{0.52\pm0.21} & o23149 \\
\formatdate{7}{7}{2023} & HD 169142 & MBI & Slow & CLC-5 & Lyot & 1.0 & 10 & 301 & 50 & \num{0.83\pm0.21} & o23166 \\
\formatdate{7}{7}{2023} & HD 191195 & MBI & - & - & - & 0.03 & 500 & 5 & $<$1 & \num{0.83\pm0.21} & o23166 \\
\formatdate{7}{7}{2023} & 21 Oph & 625-50 & - & - & - & 0.01 & 500 & 5 & $<$1 & \num{0.83\pm0.21} & o23166 \\
\formatdate{7}{7}{2023} & 21 Oph & 675-50 & - & - & - & 0.01 & 500 & 5 & $<$1 & \num{0.83\pm0.21} & o23166 \\
\formatdate{7}{7}{2023} & 21 Oph & 725-50 & - & - & - & 0.01 & 500 & 3 & $<$1 & \num{0.83\pm0.21} & o23166 \\
\formatdate{7}{7}{2023} & 21 Oph & 750-50 & - & - & - & 0.01 & 500 & 5 & $<$1 & \num{0.83\pm0.21} & o23166 \\
\formatdate{7}{7}{2023} & 21 Oph & 775-50 & - & - & - & 0.01 & 500 & 6 & $<$1 & \num{0.83\pm0.21} & o23166 \\
\formatdate{7}{7}{2023} & 21 Oph & Open & - & - & - & 0.01 & 500 & 11 & $<$1 & \num{0.83\pm0.21} & o23166 \\
\formatdate{7}{7}{2023} & HD 204827 & MBI & Fast & - & - & 0.01 & 176 & 52 & 2 & \num{0.83\pm0.21} & o23166 \\
\formatdate{7}{7}{2023} & R Aqr & H$\alpha$/H$\alpha$-Cont & - & - & - & 1.2 & 1 & 348 & 7  & \num{0.83\pm0.21} & o23166 \\
\formatdate{10}{7}{2023} & 21 Oph & MBI & - & - & - & 0.01 & 500 & 555 & 46 & \num{0.56\pm0.12} & o23166 \\
\formatdate{10}{7}{2023} & HD 163296 & MBI & Slow & CLC-3 & Lyot & 0.5 & 17 & 17 & $<$1 & \num{0.56\pm0.12} & o23166 \\
\formatdate{11}{7}{2023} & HD 1160 & MBI & - & CLC-3 & Lyot & 0.1 & 100 & 101 & 17 & \num{0.62\pm0.14} & o23166 \\
\formatdate{11}{7}{2023} & Neptune & MBI & Slow & - & - & 5 & 1 & 184 & 15 & \num{0.62\pm0.14} & o23166 \\
\formatdate{29}{7}{2023} & HIP 3373 & MBI & - & - & - & 0.1 & 50 & 50 & 4 & \num{0.74\pm0.20} & o23150 \\
\formatdate{31}{8}{2023} & HR 718 & 625-50 & - & - & - & \num{7.9e-5} & 1000 & 1 & $<$1 & \num{1.25\pm0.39} & o23013 \\
\formatdate{31}{8}{2023} & HR 718 & 675-50 & - & - & - & \num{7.9e-5} & 1000 & 1 & $<$1 & \num{1.25\pm0.39} & o23013 \\
\formatdate{31}{8}{2023} & HR 718 & 725-50 & - & - & - & \num{7.9e-5} & 1000 & 1 & $<$1 & \num{1.25\pm0.39} & o23013 \\
\formatdate{31}{8}{2023} & HR 718 & 750-50 & - & - & - & \num{7.9e-5} & 1000 & 1 & $<$1 & \num{1.25\pm0.39} & o23013 \\
\formatdate{31}{8}{2023} & HR 718 & 775-50 & - & - & - & \num{7.9e-5} & 1000 & 1 & $<$1 & \num{1.25\pm0.39} & o23013 \\
\formatdate{31}{8}{2023} & HR 718 & Open & - & - & - & \num{7.9e-5} & 1000 & 1 & $<$1 & \num{1.25\pm0.39} & o23013 \\
\formatdate{31}{8}{2023} & HR 718 & MBI & - & - & - & \num{1.1e-3} & 1000 & 1 & $<$1 & \num{1.25\pm0.39} & o23013 \\
\formatdate{31}{8}{2023} & HR 718 & H$\alpha$/H$\alpha$-Cont & - & - & - & \num{4.7e-2} & 425 & 1 & $<$1 & \num{1.25\pm0.39} & o23013 \\
\formatdate{31}{8}{2023} & HR 718 & SII/SII-Cont & - & - & - & \num{1.1e-2} & 1100 & 1 & $<$1 & \num{1.25\pm0.39} & o23013 \\
\formatdate{31}{8}{2023} & HD 19445 & H$\alpha$/H$\alpha$-Cont & - & - & - & 0.67 & 85 & 50 & 48 & \num{1.25\pm0.39} & o23013 \\
\formatdate{30}{4}{2024} & HD 137909 & 625-50 & - & - & - & \num{5.7e-4} & 500 & 10 & $<$1 & - & o24182 \\
\formatdate{30}{4}{2024} & HD 137909 & 675-50 & - & - & - & \num{5.7e-4} & 500 & 10 & $<$1 & - & o24182 \\
\formatdate{30}{4}{2024} & HD 137909 & 725-50 & - & - & - & \num{5.7e-4} & 500 & 10 & $<$1 & - & o24182 \\
\formatdate{30}{4}{2024} & HD 137909 & 750-50 & - & - & - & \num{5.7e-4} & 500 & 10 & $<$1 & - & o24182 \\
\formatdate{30}{4}{2024} & HD 137909 & 775-50 & - & - & - & \num{5.7e-4} & 500 & 10 & $<$1 & - & o24182 \\
\formatdate{30}{4}{2024} & HD 137909 & Open & - & - & - & \num{5.7e-4} & 500 & 10 & $<$1 & - & o24182 \\
\formatdate{30}{4}{2024} & HD 137909 & MBI & - & - & - & \num{1.6e-3} & 250 & 7 & $<$1 & - & o24182 \\
\formatdate{30}{4}{2024} & HD 139341 & MBI & - & - & - & \num{0.1} & 250 & 2 & $<$1 & - & o24182 \\
\enddata
\tablecomments{The number of coadds and number of frames from each dataset are the average values (not all data cubes have the same number of frames). Seeing data obtained from the MKWC archive\footnote{\url{http://mkwc.ifa.hawaii.edu/current/seeing/index.cgi}}, when available; values are the average and standard deviation DIMM measurements for the whole night.}
\end{deluxetable*}
\section{Instrument models\label{sec:models}}

\begin{figure*}
    \centering
    \includegraphics[width=\textwidth]{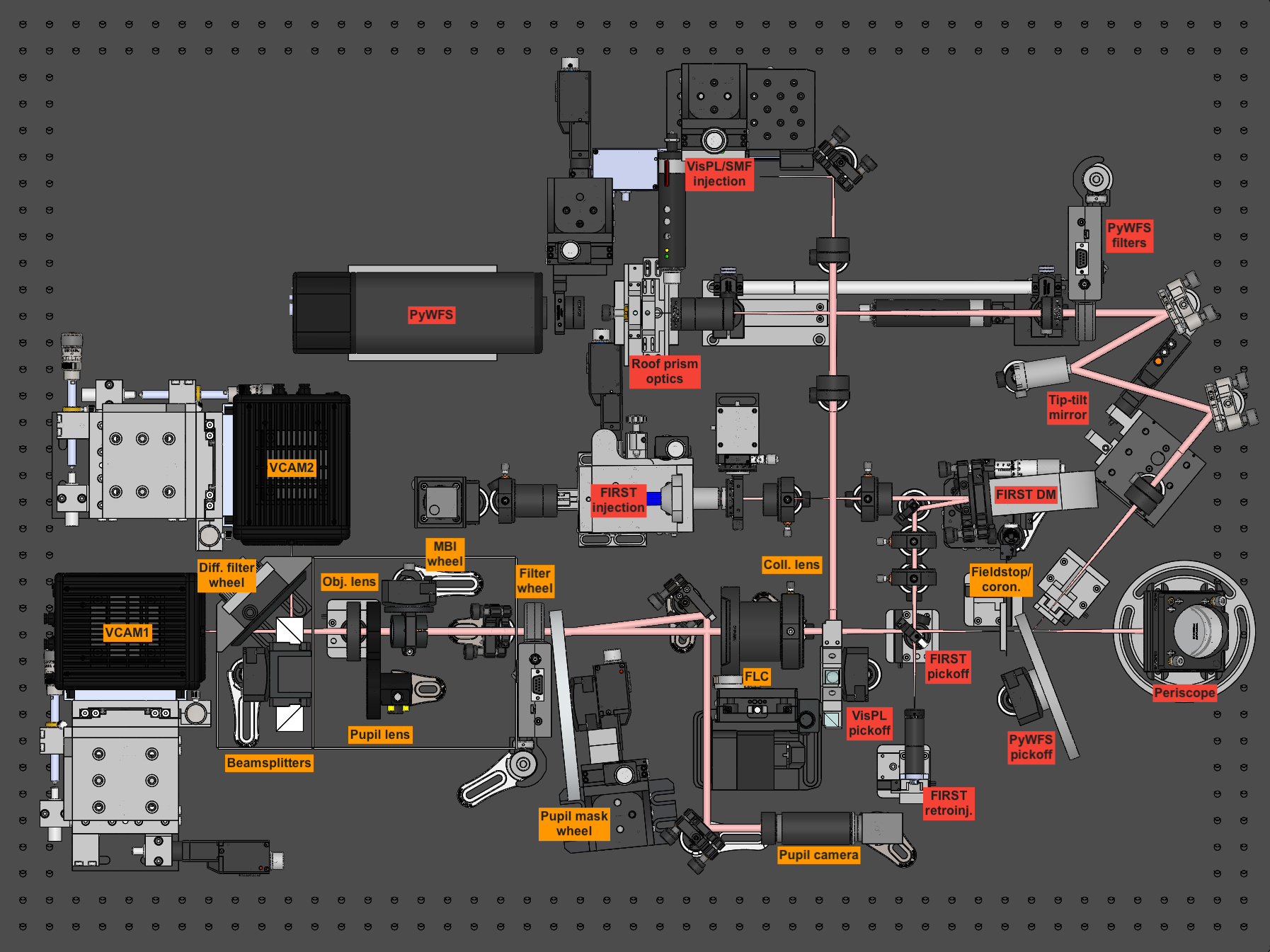}
    \caption{SCExAO visible bench including the PyWFS, FIRST injection, and VAMPIRES. VAMPIRES optics are labeled in orange and FIRST/PyWFS optics are labeled in red. The total size of the optical table shown is \SI{120}{\cm} by \SI{90}{\cm}. Parts of the baffling around VAMPIRES optics, as well as the optical bench enclosure, are hidden for clarity.\label{fig:cad}}
\end{figure*}

An annotated CAD model of VAMPIRES on the visible bench of SCExAO is shown in \autoref{fig:cad}

\section{Photon number resolving statistics}\label{sec:pnr_derivation}

We derive the theoretical noise effects of photon-number resolving data \textit{post-hoc}, wherein data that is assumed to be perfectly bias-subtracted and flat-corrected is converted to electron flux using the detector gain and rounded to the nearest whole number.

Photons that enter the detector have a certain probability of becoming photo-electrons (hereafter electrons), described by the quantum efficiency. We ignore this for now and describe the input flux in electrons. These electrons are subject to counting statistics due to their Poisson nature. Let $H$ be the rate of electrons per pixel (the quanta exposure). The probability of measuring a signal of $k$ electrons is then
\begin{equation}
    \mathcal{P}\left(k | H\right) = \frac{H^k e^{-H}}{k!}.
\end{equation}

Now consider the Gaussian read noise, $\sigma_{RN}$
\begin{equation}
    \mathcal{N}\left(x | k, \sigma_RN\right) = \frac{1}{\sigma_{RN}\sqrt{2\pi}}e^{-(x - k)^2 / 2\sigma_{RN}^2}.
\end{equation}
The probability of measuring $U$ electrons after both processes is the convolution of the two distributions
\begin{equation}
    P(U | H, \sigma_{RN}) = \sum_{k=0}^\infty{\mathcal{N}\left(U | k, \sigma_{RN}\right)\cdot \mathcal{P}\left(k|H\right)}.
\end{equation}

Qualitatively, recognize that this can be described as discrete Poisson peaks convolved with a Gaussian of width $\sigma_{RN}$. Consider the simple case of $H=0$; the data is a Gaussian centered at zero. If this data is rounded to the nearest whole number, values $|U|>0.5$ get rounded to -1 and 1, respectively, or for $|U| > 1.5$ to -2 and 2, and so on, causing variance in the measured photon number.

Consider a new distribution describing the probability of measuring values which will be rounded to the correct bins, $h(k)$
\begin{align}
    h(k) = &P\left(|U - k| < 0.5 | \sigma_{RN}\right) \\
    &= \int_{k - 0.5}^{k + 0.5}{\mathcal{N}(x | k, \sigma_{RN}) dx}.
\end{align}
This can be rewritten using the cumulative distribution of the unit normal, $\Phi$, 
\begin{equation}
    h(k) = \Phi\left(\frac{k + 0.5}{\sigma_{RN}} \right) - \Phi\left(\frac{k - 0.5}{\sigma_{RN}} \right)
\end{equation}
with
\begin{equation}
    \Phi(z) = \frac12 \left[1 - \mathrm{erf}\left(\frac{z}{\sqrt{2}}\right) \right],
\end{equation}
where $\mathrm{erf}$ is the error function.

The expected value of this distribution is 0 by convenience of symmetry
\begin{equation}
    \mathrm{E}[h] = \sum_{k=-\infty}^{\infty}{ k \cdot h(k)} = 0,
\end{equation}
and the variance of this distribution is
\begin{equation}
\label{eqn:pnr_var}
    \mathrm{Var}[h] = \sum_{k=-\infty}^{\infty}{ k^2\cdot h(k)}.
\end{equation}
For practical applications, $k$ is truncated to \num{\pm5} since the probability of observing a Gaussian event over five standard deviations away is extremely small, below $10^{-5}$, and thus $h(k)\approx0$.

Therefore, the total noise from photon number resolving is the combination of photon noise and \autoref{eqn:pnr_var}--
\begin{equation}
    \label{eqn:pnr_std}
    \sigma = \sqrt{U + \sum_{k=-\infty}^{\infty}{ k^2\cdot \left[\Phi\left(\frac{k + 0.5}{\sigma_{RN}} \right) - \Phi\left(\frac{k - 0.5}{\sigma_{RN}}\right)\right]}}.
\end{equation}

\begin{figure}
    \centering
    \script{pnr.py}
    \includegraphics[width=\columnwidth]{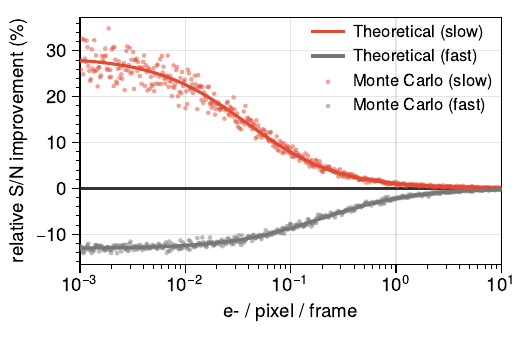}
    \caption{Relative S/N improvement from calculating the photon number via truncation in the low-flux regime compared to the standard noise terms. Solid curves are theoretical values (\autoref{eqn:pnr_std}) and the scatter points are Monte Carlo statistical simulations shown for both the ``slow'' ($\sigma_{RN}$=\SI{0.25}{\electron}) and the ``fast''  ($\sigma_{RN}$=\SI{0.4}{\electron})  detector readout modes.\label{fig:pnr}}
\end{figure}

In \autoref{fig:pnr}, the relative improvement in S/N by photon number resolving (i.e., converting the signal to electrons and rounding to the nearest whole number) is shown. We also show the standard deviation of Monte Carlo samples ($N=10^4$) with and without rounding to confirm our theoretical results. Note that there are more sophisticated algorithms for achieving higher S/N ratios with photon-number resolving than simply rounding after the fact \citep[see][]{harpsoe_bayesian_2012}, but this is still a somewhat contrived use case for a ground-based instrument like VAMPIRES, which will be photon-noise limited for most observations.

\section{Multiband Imaging Ghosts}\label{sec:ghosts}

When stacking dichroics with low angles of incidence, extra reflections create ghost PSFs. Our multiband dichroic technique (\autoref{sec:mbi}) uses three dichroics in front of a mirror with $\sim$\ang{0.4} tilts between each filter with respect to the mirror. We characterized the ghost PSFs using a dispersing prism to understand their origin by analyzing spectra.
\begin{figure}
    \centering
    \script{mbi_ghosts_diagram.py}
    \includegraphics[width=0.85\columnwidth]{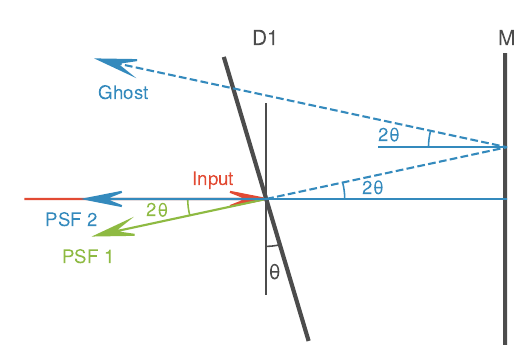}
    \caption{Raytrace diagram for ghosts from stacked dichroics. D1 represents a dichroic tilted at angle $\theta$, and M represents a standard mirror. The incoming ray (red arrow) goes from left to right. The light reflected by D1 is shown with a green arrow. Light transmitted by D1 is blue-- the solid line is the main ray and the dashed line is the first ghost ray.\label{fig:mbi_ghost}}
\end{figure}

To explain, consider the simple case of a single dichroic with a tilt, $\theta$, in front of a mirror (\autoref{fig:mbi_ghost}). There are three rays to explain the ghost behavior-- the first is the light reflected by the dichroic (green arrow), which has an angle of $-2\theta$. The second ray is transmitted through the dichroic, reflects off the mirror, and returns with an angle of \ang{0} (solid blue arrow). Because the dichroic has parallel faces, there is no angular deviation of the transmitted rays. The angle between the solid blue and green arrows ($2\theta$) is the characteristic separation between the fields in the focal plane.

The final ray is formed from a transmitted beam that reflects off the rear mirror and then off the dichroic's backside (dashed blue arrow). The reflection is a natural consequence of the dichroic filter curve-- any light that is not fully transmitted will be reflected by the dielectric coating and return towards the mirror, picking up an angle of $2\theta$ due to the tilt of the dichroic. This beam reflects off the rear mirror again and then is transmitted through the dichroic with a ray angle of $2\theta$. This will form a PSF with the same spectrum as ``PSF 2'' but separated by an angle of $2\theta$. More back reflections can occur every time the light transmits through the dichroic, reflecting a small fraction of the light at integer multiple separations of $2\theta$ (i.e., $4\theta$, $6\theta$, and so on). These higher-order ghosts are less concerning because they are exponentially fainter and far from the primary PSFs.

One way to remove these ghosts would be to improve the dielectric coatings of the filters until the ghosts are undetectable. Our multiband dichroics have a transmission of $\sim$95\% outside of the OD=$4$ blocking region, meaning $\sim$5\% of light is reflected into these ghosts and easily detected. It is difficult and expensive to manufacture dichroics that simultaneously have efficient blocking (OD$>$3) in band and high throughput out of band ($>$99.9\%), so we opted to address the ghosts geometrically-- creating a focal plane layout where no ghosts overlap with any science fields.

\begin{figure}
    \centering
    \script{mbi_field_diagram.py}
    \includegraphics[width=\columnwidth]{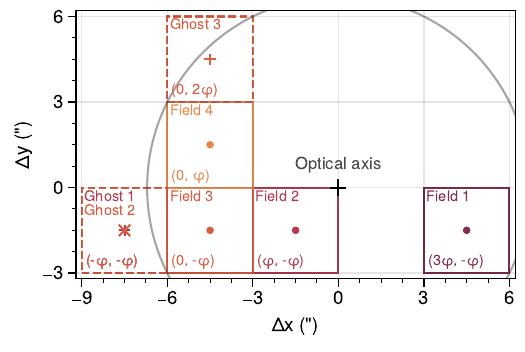}
    \caption{Diagram of the fields produced by the VAMPIRES multiband dichroics. Each field is colored based on the expected spectrum, from brown (F760) to orange (F610). Each field is labeled in the bottom left with the angular displacement ($x,y$) of the field in terms of $\varphi=2\theta$, the magnified angle offset between dichroic filters in the MBI stack.\label{fig:mbi_fields}}
\end{figure}

Our solution uses optimal angles for the dichroic tilts which place any ghost PSFs (and their \ang{;;3} FOV) completely out of the FOV of the primary fields (\autoref{fig:mbi_fields}). Our design uses as few pixel rows as possible to maximize detector readout speed. Additionally, a single field can be cropped out to double the readout speed leaving three fields remaining. For the final design, angles were chosen for each dichroic slightly larger than \ang{;;3} so that the manufacturing tolerances would allow for at least $\sim$\SI{10}{\pixel} between fields.

\section{Measuring the Non-telecentricity of VAMPIRES}\label{sec:telecentricity}

\begin{figure}
    \centering
    \script{pinhole_defocus.py}
    \includegraphics[width=\columnwidth]{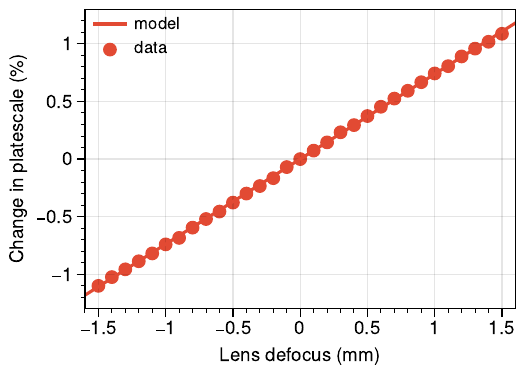}
    \caption{The percentage change in plate scale as the image passes through focus due to the non-telecentricity of the optical system. A linear model is fit to the data and shown with a solid line.\label{fig:defocusing}}
\end{figure}

VAMPIRES is non-telecentric, which means that a shift in focus affects the plate scale in the detector plane. Given the instrument's space constraints, this is nearly impossible to avoid optically. In theory, if the instrument F-ratio is stable and the detector is reasonably well-focused, then the plate scale should stay stable. Nonetheless, we quantify the change in plate scale as a function of defocus using the SCExAO internal pinhole mask.

For this experiment, we used the pinhole mask and moved the detector stage in steps around the optimal focus. For each frame, PSFs were fit to the pinhole grid, and the average separation was calculated. The relative displacement of the grid as a function of defocus is shown in \autoref{fig:defocusing}.

These data were fit with a linear model correlating the percentage change in plate scale to the absolute defocus in \si{\milli\meter}. The fit is shown in \autoref{fig:defocusing} which has a slope of \SI{1.106(3)}{\%/\milli\meter}. This means the objective lens must stay focused to better than \SI{0.09}{\milli\meter} precision to keep the astrometric precision below 0.1\%, which is not difficult to achieve when focusing with the internal laser.

\section{Data and Code Availability}
This study used the reproducibility software \texttt{showyourwork} \citep{luger_mapping_2021}, which leverages continuous integration to create the figures and compile this manuscript programmatically. Each figure caption contains a link to the script used to make the figure (at the commit corresponding to the current build of the manuscript). The git repository associated with this study is publicly available at \url{https://github.com/mileslucas/vampires_instrument_paper}. Reduced data is available upon request, and raw observational data is available from the SMOKA archive\footnote{\url{https://smoka.nao.ac.jp/}}.

\end{document}